\documentclass[preprint,12pt,authoryear]{elsarticle}

\makeatletter
\def\ps@pprintTitle{%
   \let\@oddhead\@empty
   \let\@evenhead\@empty
   \def\@oddfoot{\reset@font\hfil\thepage\hfil}
   \let\@evenfoot\@oddfoot
}
\makeatother

\usepackage{latexsym,bm}
\usepackage{amssymb,amsmath,amsfonts}
\usepackage[usenames]{color}
\usepackage{graphicx}
\usepackage{url}
\usepackage[markup=default]{changes}
\usepackage{booktabs}
\usepackage{subcaption}
\usepackage{xr}
\usepackage{ulem}
\begin{document}

\begin{frontmatter}

\title{Spatially Adaptive Calibrations of AirBox PM$_{2.5}$ Data}

\author[nsysu]{ShengLi Tzeng}
\author[nthu]{Chi-Wei Lai}
\author[as]{Hsin-Cheng Huang\corref{cor1}}
  \ead{hchuang@stat.sinica.edu.tw}

\address[nsysu]{Department of Applied Mathematics, National Sun Yat-sen University, Taiwan, R.O.C.}
\address[nthu]{Institute of Statistics, National Tsing Hua University, Taiwan, R.O.C.}
\address[as]{Institute of Statistical Science, Academia Sinica, Taiwan, R.O.C.}

\cortext[cor1]{Corresponding author}

\date{}

\begin{abstract}
\baselineskip=18pt
Two networks are available to monitor PM$_{2.5}$ in Taiwan,
including the Taiwan Air Quality Monitoring Network (TAQMN) and the AirBox network.
The TAQMN, managed by Taiwan's Environmental Protection Administration (EPA), provides high-quality PM$_{2.5}$ measurements at $77$ monitoring stations.
More recently, the AirBox network was launched, consisting of low-cost, small internet-of-things (IoT) microsensors (i.e., AirBoxes) at thousands of locations.
While the AirBox network provides broad spatial coverage, its measurements are not reliable and require calibrations.
However, applying a universal calibration procedure to all AirBoxes does not work well
because the calibration curves vary with several factors, including the chemical compositions of PM$_{2.5}$, which are not homogeneous in space.
Therefore, different calibrations are needed at different locations with different local environments.
Unfortunately, most AirBoxes are not close to EPA stations, making the calibration task challenging.
In this article, we propose a spatial model with spatially varying coefficients to account for heteroscedasticity in the data.
Our method gives adaptive calibrations of AirBoxes according to their local conditions
and provides accurate PM$_{2.5}$ concentrations at any location in Taiwan, incorporating two types of measurements.
In addition, the proposed method automatically calibrates measurements from a new AirBox once it is added to the network.
We illustrate our approach using hourly PM$_{2.5}$ data in the year 2020.
After the calibration, the results show that the PM$_{2.5}$ prediction improves about 37\% to 67\% in root mean-squared prediction error
for matching EPA data.
In particular, once the calibration curves are established,
we can obtain reliable PM$_{2.5}$ values at any location in Taiwan, even if we ignore EPA data.
\bigskip
\end{abstract}

\begin{keyword}
\baselineskip=18pt
Heterogeneous variance; kriging; microsensor; monitoring station; robust estimation; spatially varying coefficient model.
\end{keyword}

\end{frontmatter}

\baselineskip=22.5pt

\section{Introduction}

Two networks composed of two different types of instruments are available in monitoring the PM$_{2.5}$ process in Taiwan.
Traditionally, PM$_{2.5}$ is monitored by large monitoring stations
in the Taiwan Air Quality Monitoring Network (TAQMN) of the Environmental Protection Administration (EPA).
These monitoring stations provide high-quality hourly measurements, but are costly to establish and operate.
Currently, TAQMN consists of $77$ stations (74 on the main island and 3 on the offshore islands),
which are deployed at heights of approximately 10 meters above the ground.
Recently, another network consisting of small, low-cost, internet-of-things microsensors, called AirBoxes, is established.
Although measurements from AirBoxes based on optical sensors are not as accurate as those from EPA monitoring stations,
they are broadly deployed (at around 3 meters height) and generate data about every 5 minutes, resulting in high spatial and temporal coverage.

\begin{figure}[tb]\centering
\begin{tabular}{cc}
\includegraphics[scale=0.32,trim={0cm 0cm 0cm 0cm},clip]{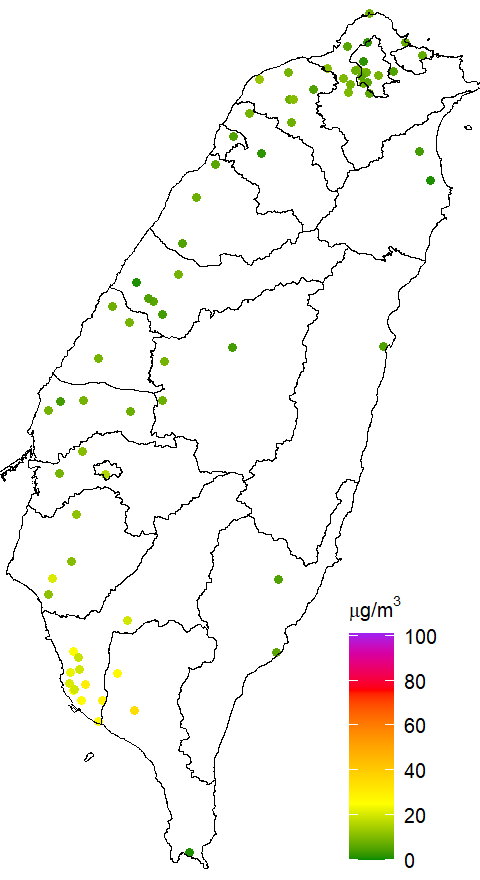}
&  \includegraphics[scale=0.32,trim={0cm 0cm 0cm 0cm},clip]{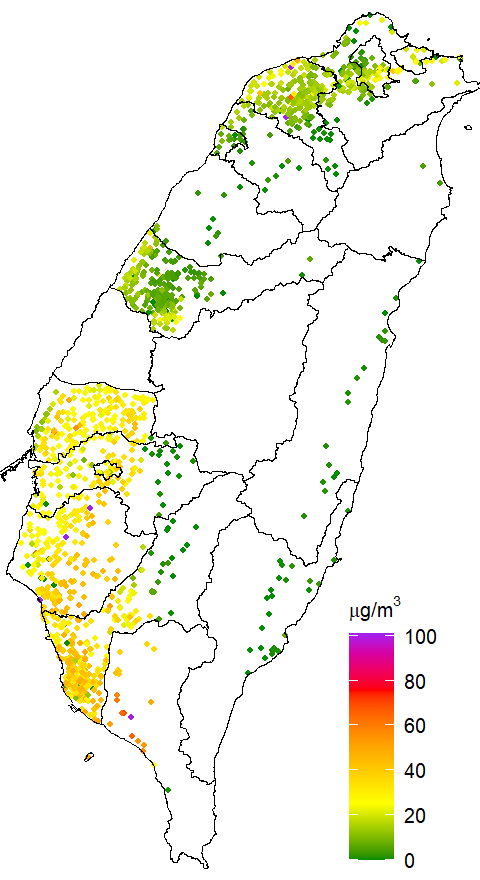}\\

(a) & (b)
\end{tabular}
\caption{PM$_{2.5}$ measurements at 9:00am on December 22, 2020: (a) from 74 EPA stations;
(b) from 1769 AirBoxes.}
\label{fig:data}
\end{figure}

Figures~\ref{fig:data}(a) and \ref{fig:data}(b) show the EPA data (in ppm from 74 stations on the main island)
and the AirBox data (in ppm from 1769 AirBoxes) at 9:00am on December 22, 2020.
Clearly, the AirBox network is considerably denser.
However, as seen in Figures~\ref{fig:data}(a) and \ref{fig:data}(b),
measurements from AirBoxes have higher variances and are usually higher than those from EPA stations.
These upward biases are partly caused by higher altitudes of EPA stations than AirBoxes,
as PM$_{2.5}$ concentrations tend to be lower at higher altitudes.
There are also some clear outlying measurements.
The goal of this article is to develop a reliable calibration method
so that calibrated PM$_{2.5}$ measurements from AirBoxes are consistent with those from EPA stations.

One effective way to calibrate AirBoxes is by regression, which works well for AirBoxes colocated at (or very close to) EPA stations.
The method is particularly effective if it suffices to apply a universal calibration line to all AirBoxes.
However, as shown in the next section, different calibrations are needed for AirBoxes located at various locations
with different local environments.
Additionally, most AirBoxes are away from EPA stations and cannot be calibrated directly, challenging the calibration task.
We propose to explore the relationship in the two datasets
and leverage the proximity of nearby AirBoxes and EPA stations using a spatially varying-coefficients model.

The rest of this paper is organized as follows. Section 2 compares PM$_{2.5}$ data between the two networks.
In Section 3, we introduce our proposed calibration method.
We also provide a robust parameter estimation procedure and a spatial prediction method that incorporates the two types of measurements.
The calibration results are given in Section 4. Finally, Section 5 provides a brief summary.

\section{Comparisons between EPA and AirBox Data}

We compare EPA and AirBox data observed in December, 2020 to give some ideas about how they differ.
The data being analyzed in this paper are available at Civil IoT Taiwan Data Service Platform.
The EPA TAQMN data can be downloaded from \url{https://ci.taiwan.gov.tw/dsp/en/environmental_air_epa_en.aspx},
and the AirBox data can be downloaded from \url{https://ci.taiwan.gov.tw/dsp/history/iis_airbox/}.

Since TAQMN produces hourly data,
we first aggregate the raw AirBox data into hourly data by averaging over all data for each hour and each site. 
Thus, we obtain $\bm{z}_t=(z_t(\bm{s}_1),\dots,z_t(\bm{s}_n))'$ at locations $\bm{s}_1,\dots,\bm{s}_n\in D$ and hour $t$ with possible missing values,
where $n=$2640, $t=1,\dots,T$, and $D\in\mathbb{R}^2$ is a region containing the main island of Taiwan.
As a preliminary data analysis, we choose data at $T=666$ hours, which have at least 500 non-missing observations in each $\bm{z}_t$; $t=1,\dots,T$.
We denote the corresponding EPA data at hour $t$ by $\bm{z}^*_t=(z^*_t(\bm{s}^*_1),\dots,z^*_t(\bm{s}^*_m))'$; $t=1,\dots,T$,
where $\bm{s}^*_1,\dots,\bm{s}^*_m\in D$ and $m=74$.

\begin{figure}[tbh]\centering
\!\!\includegraphics[scale=0.245,trim={0cm 0cm 0cm 0cm},clip]{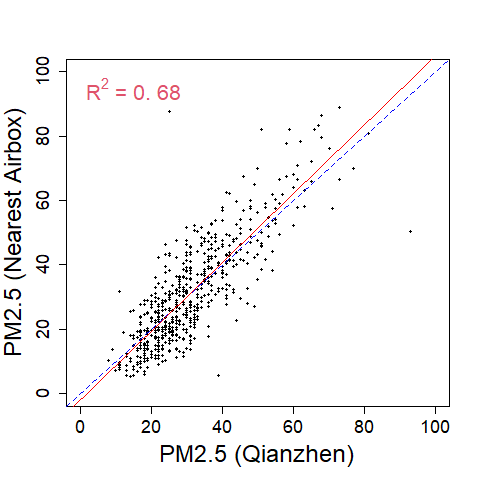}
\!\!\includegraphics[scale=0.245,trim={0cm 0cm 0cm 0cm},clip]{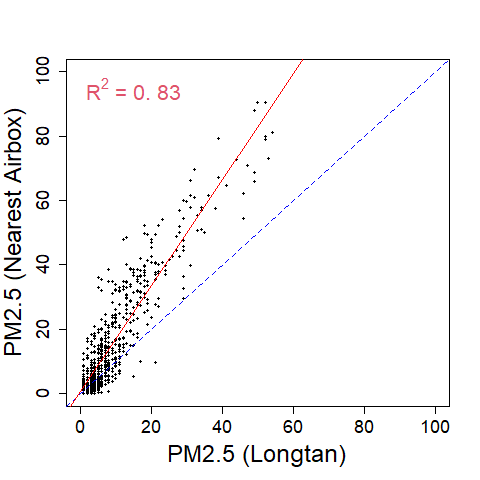}
\!\!\includegraphics[scale=0.245,trim={0cm 0cm 0cm 0cm},clip]{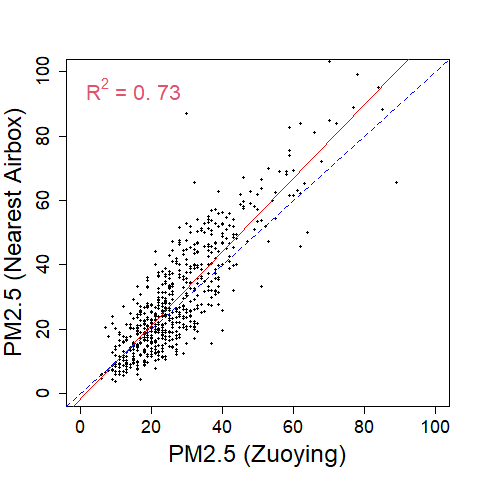}
\!\!\includegraphics[scale=0.245,trim={0cm 0cm 0cm 0cm},clip]{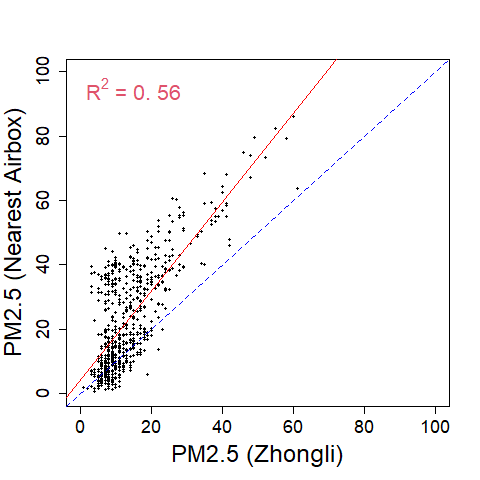}\\
\!\!\includegraphics[scale=0.245,trim={0cm 0cm 0cm 0cm},clip]{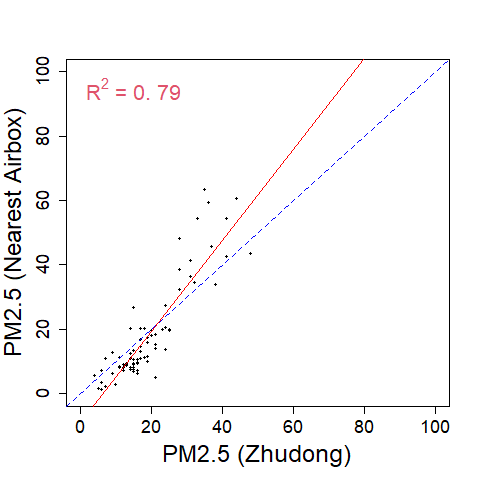}
\!\!\includegraphics[scale=0.245,trim={0cm 0cm 0cm 0cm},clip]{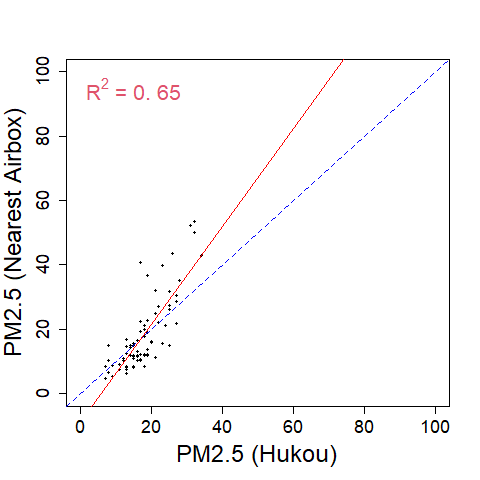}
\!\!\includegraphics[scale=0.245,trim={0cm 0cm 0cm 0cm},clip]{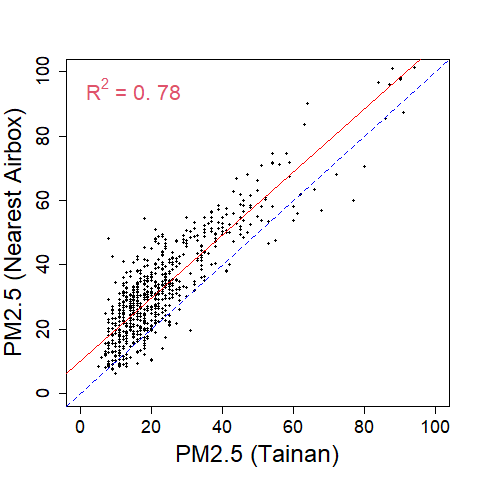}
\!\!\includegraphics[scale=0.245,trim={0cm 0cm 0cm 0cm},clip]{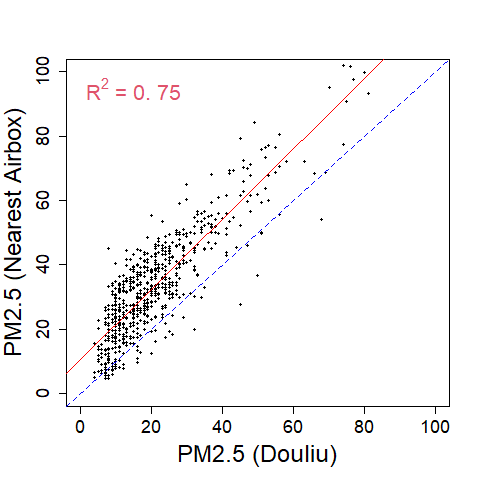}\\
\!\!\includegraphics[scale=0.245,trim={0cm 0cm 0cm 0cm},clip]{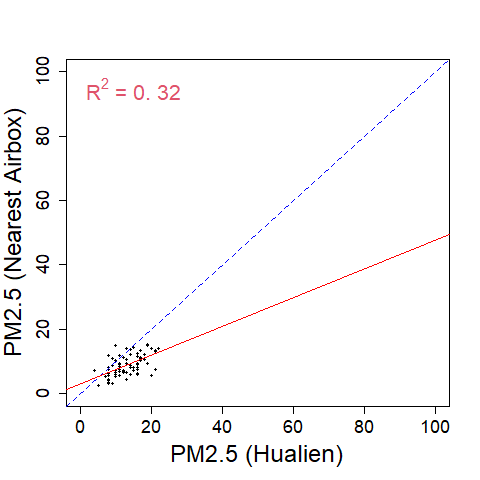}
\!\!\includegraphics[scale=0.245,trim={0cm 0cm 0cm 0cm},clip]{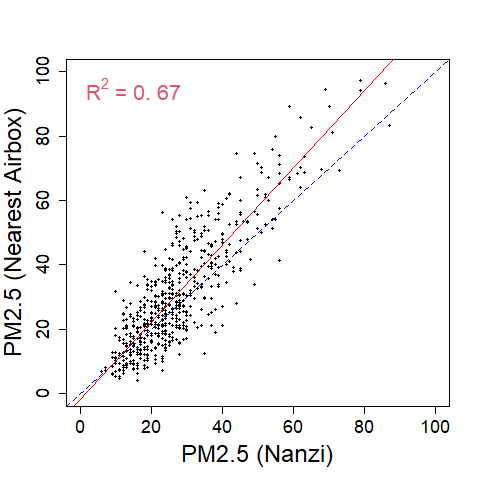}
\!\!\includegraphics[scale=0.245,trim={0cm 0cm 0cm 0cm},clip]{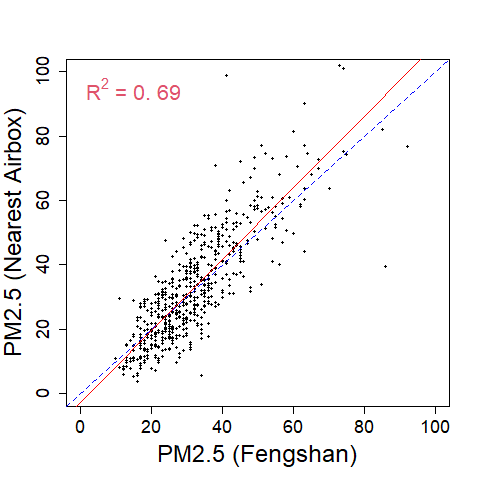}
\!\!\includegraphics[scale=0.245,trim={0cm 0cm 0cm 0cm},clip]{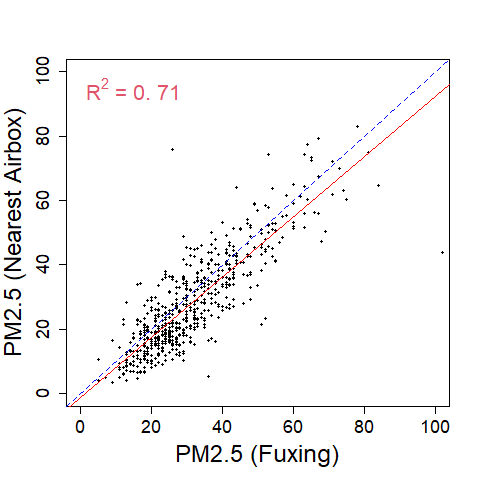}
\caption{Scatter plots of hourly PM$_{2.5}$ concentrations (in ppm) for twelve EPA stations
(in the $x$ axis with their locations shown in Figure \ref{fig:lsfit}) and
their nearest AirBoxes (in the $y$ axis) based on data in December, 2020,
where the blue dash line is the 45-degree line and the red solid line is the fitted regression line in each plot.}
\label{fig:nearest}
\end{figure}

It is known that AirBox data tend to have high variation,
and produce slightly higher PM$_{2.5}$ measurements than the corresponding EPA measurements.
This upward bias can be seen in Figures~\ref{fig:data}(a) and \ref{fig:data}(b),
and is partly caused by lower altitudes of AirBoxes (mostly deployed at around 3 meters) than EPA
equipments (placed at around 10 meters).
To examine the data more closely from the two different networks, we find 12 EPA stations that have at least 5 AirBoxes within their 2 km range.
Specifically, let $\bm{s}^*_j$ be the location of such an EPA station, then for $N^*_j\equiv\{i:\|\bm{s}_i-\bm{s}^*_j\|\leq 2\}$, we have $|N^*_j|\geq 5$.
Let $\bm{s}_{j'}\in N^*_j$ be the AirBox location nearest to $\bm{s}^*_j$. That is, $j'\equiv\mathop{\arg\min}_{i\in N^*_j}\|\bm{s}_i-\bm{s}^*_j\|$.
Figure~\ref{fig:nearest} shows scatter plots $\big\{\big(z^*_t(\bm{s}^*_{j}),z_t(\bm{s}_{j'})\big):t=1,\dots,T\big\}$ of hourly PM$_{2.5}$ (in ppm)
between the two subsets of data at the 12 EPA stations.
Although $\big\{z^*_t(\bm{s}^*_j):t=1,\dots,T\big\}$ and $\big\{z_t(\bm{s}_{j'}):t=1,\dots,T\big\}$ are positively correlated,
their coefficient-of-determination ($R^2$) values ranging only from $0.32$ to $0.83$.
The discrepancies between AirBox and EPA data have made AirBoxes a significant concern regarding their measuring accuracy.

\begin{figure}[tbh]\centering
\!\!\includegraphics[scale=0.245,trim={0cm 0cm 0cm 0cm},clip]{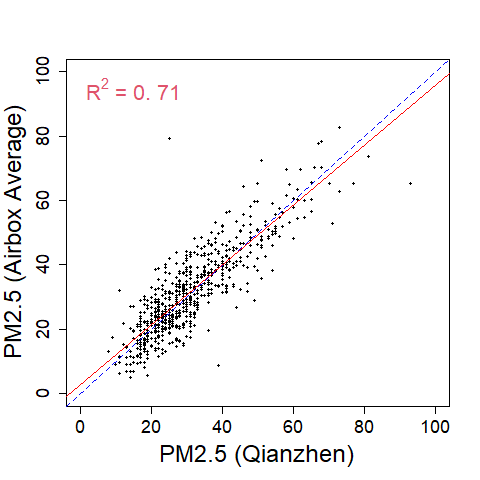}
\!\!\includegraphics[scale=0.245,trim={0cm 0cm 0cm 0cm},clip]{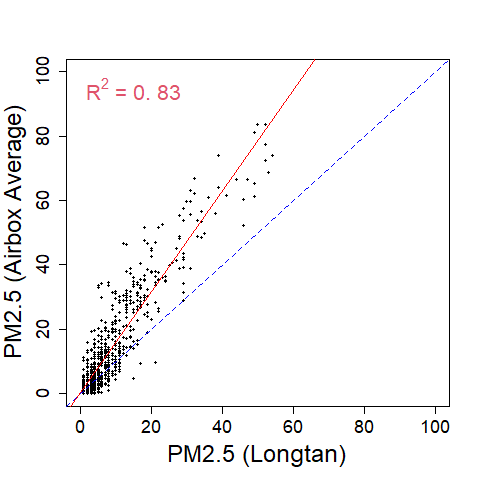}
\!\!\includegraphics[scale=0.245,trim={0cm 0cm 0cm 0cm},clip]{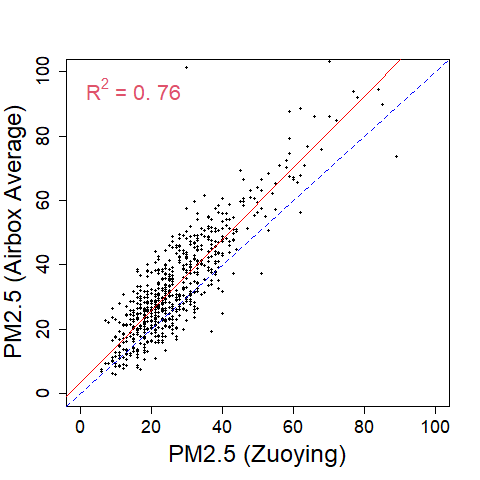}
\!\!\includegraphics[scale=0.245,trim={0cm 0cm 0cm 0cm},clip]{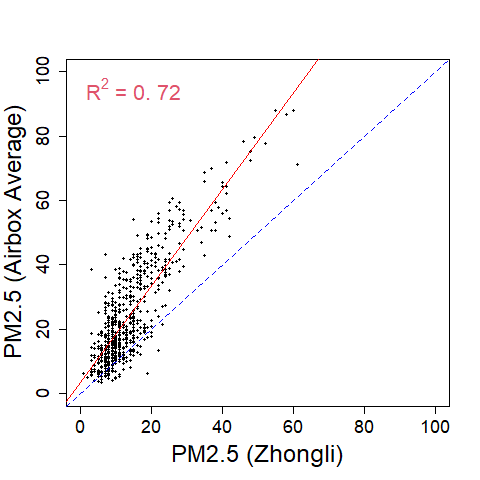}\\
\!\!\includegraphics[scale=0.245,trim={0cm 0cm 0cm 0cm},clip]{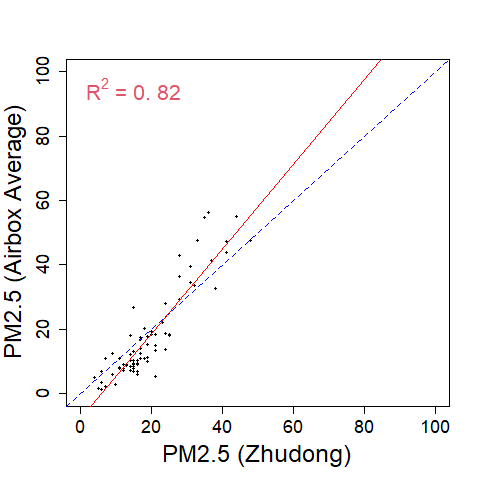}
\!\!\includegraphics[scale=0.245,trim={0cm 0cm 0cm 0cm},clip]{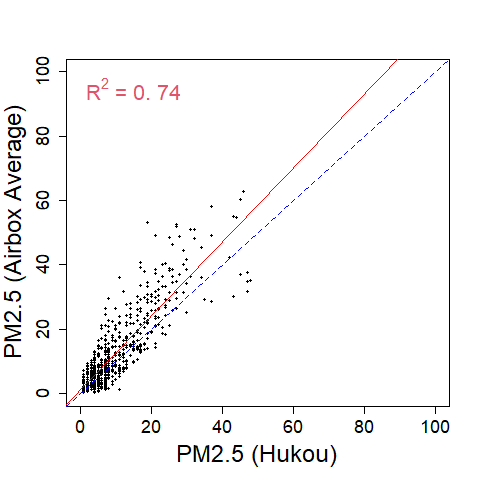}
\!\!\includegraphics[scale=0.245,trim={0cm 0cm 0cm 0cm},clip]{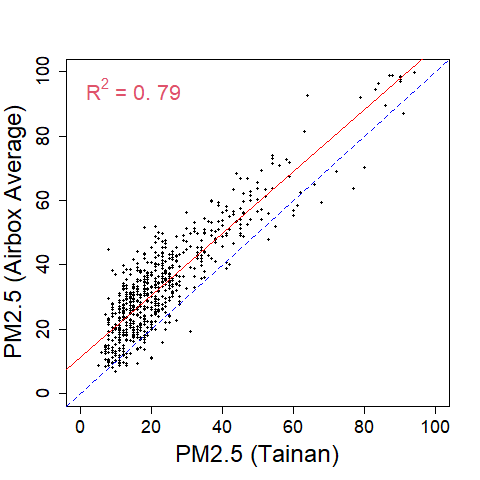}
\!\!\includegraphics[scale=0.245,trim={0cm 0cm 0cm 0cm},clip]{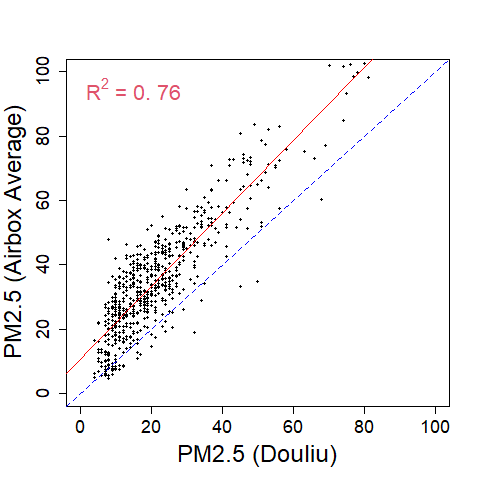}\\
\!\!\includegraphics[scale=0.245,trim={0cm 0cm 0cm 0cm},clip]{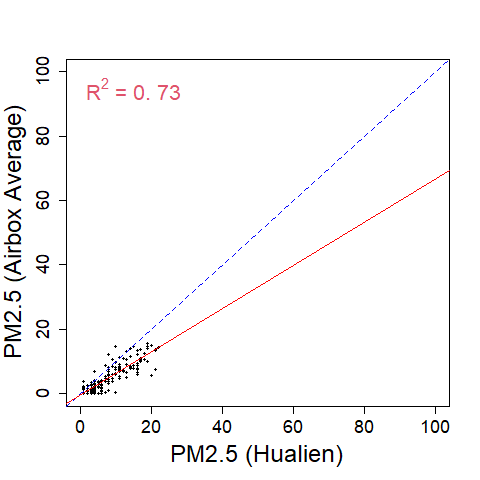}
\!\!\includegraphics[scale=0.245,trim={0cm 0cm 0cm 0cm},clip]{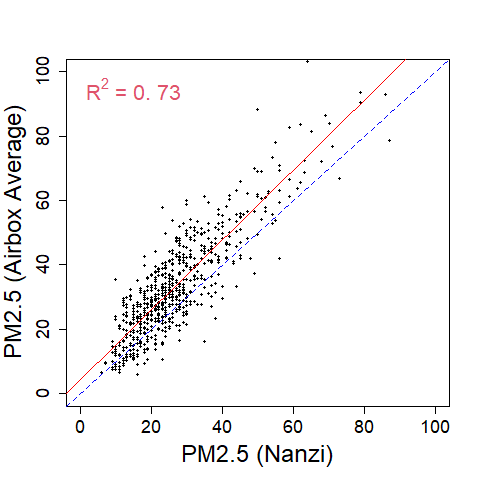}
\!\!\includegraphics[scale=0.245,trim={0cm 0cm 0cm 0cm},clip]{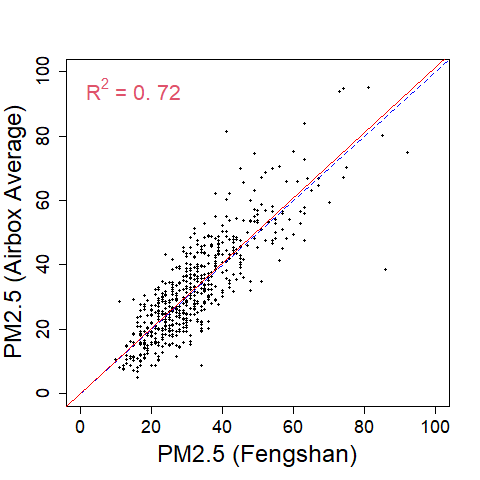}
\!\!\includegraphics[scale=0.245,trim={0cm 0cm 0cm 0cm},clip]{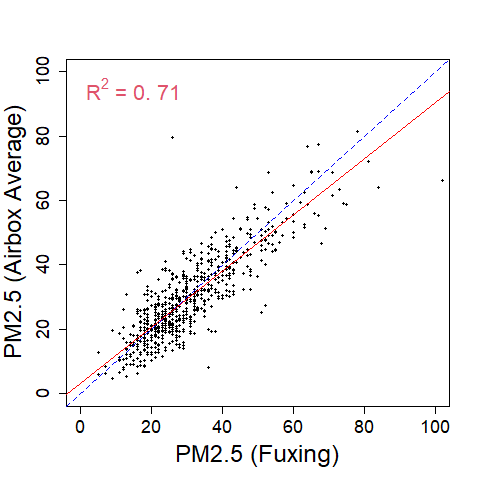}
\caption{Scatter plots of hourly PM$_{2.5}$ concentrations (in ppm) for twelve EPA stations
(in the $x$ axis with their locations shown in Figure \ref{fig:lsfit}) and
the average PM$_{2.5}$ values (in ppm) over the corresponding AirBoxes within 2 km radius (in the $y$ axis)
based on data in December, 2020, where the blue dash line is the 45-degree line and the red solid line is the fitted regression line in each plot.}
\label{fig:ave}
\end{figure}

Nevertheless, it is possible to reduce the variance by utilizing many closely located AirBoxes.
We take a simple average over the AirBox measurements falling within 2 km radius of an EPA station at each hour $t$.
Substituting these for the corresponding AirBox measurements in Figure~\ref{fig:nearest}, we obtain Figure~\ref{fig:ave}
with each plot showing points $\big\{\big(z^*_t(\bm{s}^*_{j}),\bar{z}_t(\bm{s}^*_j)\big):t=1,\dots,T\big\}$ corresponding to an EPA station at $\bm{s}^*_j$,
where $\bar{z}_t(\bm{s}^*_j)\equiv\displaystyle\frac{1}{|N^*_j|}\sum_{i\in N^*_j}z_t(\bm{s}_i)$.
Comparing Figure~\ref{fig:ave} with Figure~\ref{fig:nearest},
the averaged PM$_{2.5}$ values from AirBoxes can be seen to match the EPA measurements much better with larger $R^2$ values.
Indeed, unity is strength; even though one AirBox is not very useful, many together can average out noise.
However, the intercepts and the slopes of the fitted regression lines appear to vary from sites to sites
with no common pattern, showing the need of site dependent calibrations.

\begin{figure}[tbh]\centering
\begin{tabular}{cc}
  \includegraphics[scale=0.32,trim={0cm 0cm 0cm 0cm},clip]{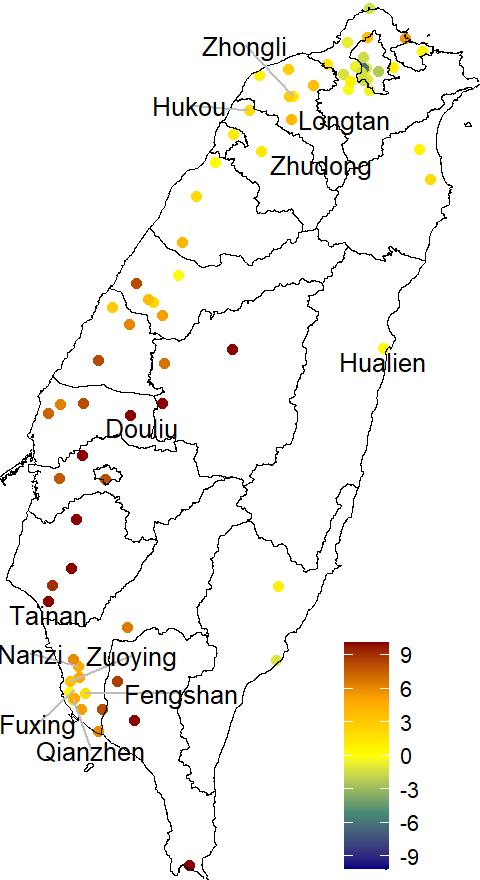}
&  \includegraphics[scale=0.32,trim={0cm 0cm 0cm 0cm},clip]{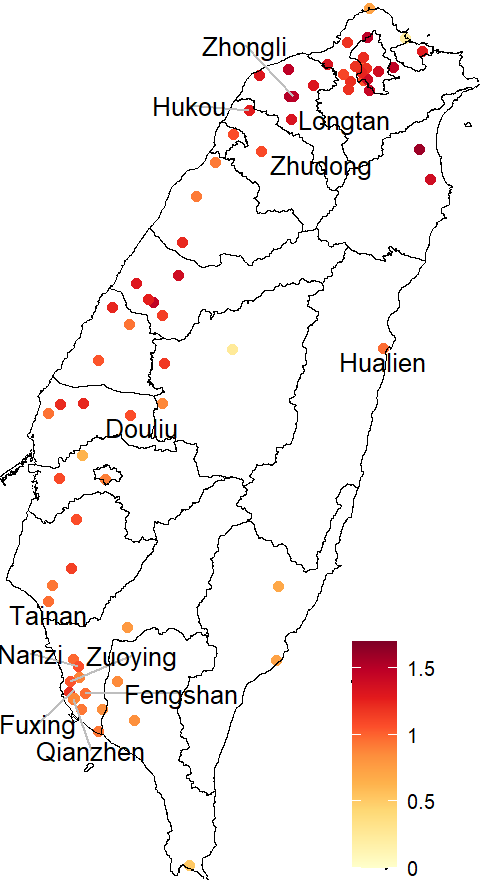}\\
(a) & (b)
\end{tabular}
\caption{Maps of (a) intercepts and (b) slopes of the preliminary calibrated regression lines obtained
by regressing kriged AirBox predicted values on the corresponding EPA data.}
\label{fig:lsfit}
\end{figure}

To visualize how these regression lines vary in space,
we first perform ordinary kriging to obtain a PM$_{2.5}$ predicted surface
for each $t=1,\dots,T$ based on AirBox data $\bm{z}_t$
using the isotropic exponential covariance model with the nugget effect,
where the parameters are estimated by maximum likelihood.
Let $\{\tilde{z}^{(\mathrm{ok})}_t(\bm{s}):\bm{s}\in D\}$ be the ordinary-kriging surface, for $t=1,\dots,T$.
Then for each $j=1,\dots,74$, we regress $\tilde{z}^{(\mathrm{ok})}_t(\bm{s}^*_j)$ on $z^*_t(\bm{s}^*_j)$ based on
$\big\{\big(z^*_t(\bm{s}^*_j),\tilde{z}^{(\mathrm{ok})}_t(\bm{s}^*_j)\big):t=1,\dots,T\big\}$ and obtain 74 calibrated regression lines
corresponding to 74 EPA stations in the main island Taiwan.
The resulting intercepts and slopes are illustrated as maps in Figure~\ref{fig:lsfit}.
Surprisingly, both the intercepts and the slopes exhibit spatial patterns.
For example, the intercepts are smoothly varying and mainly positive, with larger values in the south than in the north.
In contrast, the slopes are mostly less than one in the south but greater than one in the north.
These patterns are likely caused by similar chemical compositions of PM$_{2.5}$ at nearby locations,
which motivates us to develop a statistical model accounting for the patterns.

\section{The Proposed Method}

\subsection{The calibration model}
\label{sec:model}

Let $\{y^*_t(\bm{s}):\bm{s}\in D\}$ be the underlying PM$_{2.5}$ process at time $t$, which is assumed to be a Gaussian spatial process.
We observe two types of data, which are given by the following measurement equations, for $t=1,\dots,T$:
\begin{align}
   z_t(\bm{s}_i)
=&~ y_t(\bm{s}_i)+\varepsilon_t(\bm{s}_i);\quad i=1,\dots,n,
\label{eq:data1}\\
  z^*_t(\bm{s}^*_j)
=&~ y^*_t(\bm{s}^*_j)+\xi_t(\bm{s}^*_j);\quad j=1,\dots,m,
\label{eq:data2}
\end{align}	

\noindent where $\{y_t(\bm{s}):\bm{s}\in D\}$ is a hidden Gaussian process,
$\varepsilon_t(\bm{s}_i)\sim N(0,\sigma_{\varepsilon t}^2(\bm{s}_i))$; $i=1,\dots,n$,
and $\xi_t(\bm{s}^*_j)\sim N(0,\sigma_{\xi t}^2)$; $j=1,\dots,m$.
Here $\{\varepsilon_t(\bm{s}_i)\}$ and $\{\xi_t(\bm{s}^*_j)\}$ are mutually uncorrelated.
Our goal is to find a calibration formula for $z_t(\cdot)$
so that after calibration it is close to $y^*_t(\cdot)$ as much as possible, for $t=1,\dots,T$.

Since we expect $\sigma_{\xi t}^2$ to be very small, we assume it to be zero in our data analysis.
That is, $z^*_t(\bm{s}^*_j)$ measures PM$_{2.5}$ precisely with $z^*_t(\bm{s}^*_j)=y^*_t(\bm{s}^*_j)$,
for $j=1,\dots,m$ and $t=1,\dots,T$.
On the other hand,  $\sigma_{\varepsilon t}^2(\bm{s}_i)$ tends to increase with $y_t(\bm{s}_i)$.
We model it as a piecewise linear function of $y(\bm{s}_i)$ with two pieces (having slopes $a_1$ and $a_2$) connected at $a_3$:
\begin{equation}
\sigma^2_{\varepsilon t}(\bm{s}_i)=a_0+a_1\, y_t(\bm{s}_i)+(a_2-a_1)(y_t(\bm{s}_i)-a_3)_+;\quad i=1,\dots,n,\,t=1,\dots,T,
\label{eq:sigmasq}
\end{equation}

\noindent where $\bm{a}=(a_0,a_1,a_2,a_3)'\in[0,\infty)^4$ consists of unknown parameters and $c_+\equiv\max(c, 0)$.

Motivated by the preliminary calibration results at EPA locations demonstrated on Figure~\ref{fig:lsfit},
we propose the following calibration equation, which links $y(\bm{s})$ to the true PM$_{2.5}$ process $y^*(\bm{s})$:
\begin{align}
y^*(\bm{s}) =&~ f_0(\bm{s})+f_1(\bm{s})y(\bm{s});\quad\bm{s}\in D,
\label{eq:calibration}
\end{align}

\noindent where $f_0(\cdot)$ is an unknown function of the intercept and $f_1(\cdot)$ is an unknown function of the slope.
Following Huang \textit{et al}.~(2018), we model the hidden Gaussian process $y_t(\cdot)$ corresponding to AirBoxes as:
\begin{align}
  y_t(\bm{s})
=&~\bm{\varphi}(\bm{s})'\bm{\alpha}_t+\bm{x}_t(\bm{s})'\bm{\beta}_t+\eta_t(\bm{s})\notag\\
=&~ \sum_{k=1}^{K}\alpha_k\varphi_k(\bm{s})+\sum_{\ell=1}^p\beta_{t\ell}x_{t\ell}(\bm{s})+\eta_t(\bm{s});\quad \bm{s}\in D,\,t=1,\dots,T,
\label{eq:model}
\end{align}

\noindent where $\mathrm{E}(y_t(\bm{s}))=\bm{\varphi}(\bm{s})'\bm{\alpha}_t+\bm{x}_t(\bm{s})'\bm{\beta}_t$ captures the large-scale features in terms of regressors,
$\bm{\varphi}(\bm{s})=(\varphi_1(\bm{s}),\dots,\varphi_{K}(\bm{s}))'$ are the first $K$ multi-resolution spline basis functions
with respect to the control points $\{\bm{s}_1,\dots,\bm{s}_n\}$
proposed by Tzeng and Huang (2018),
$\bm{x}_t(\bm{s})=(x_{t1}(\bm{s}),\dots,x_{tp}(\bm{s}))'$ consists of $p$ covariates,
$(\bm{\alpha}'_t,\bm{\beta}'_t)'\in\mathbb{R}^{K+p}$ are regression coefficients,
and $\eta_t(\cdot)$ is a zero-mean spatial dependent process parametrized by the isotropic exponential covariance model:
\[
C_t(\bm{s}-\bm{u})\equiv\mathrm{cov}(\eta_t(\bm{s}),\eta_t(\bm{u}))=v^2_t\exp(-\|\bm{s}-\bm{u}\|/\lambda_t);\quad\bm{s},\bm{u}\in D,\,t=1,\dots,T,
\]
with $\{v^2_t\}$ and $\{\lambda_t\}$ being the variance and the range parameters.

\subsection{Parameter estimation}
\label{sec:estimation}

We assume that $\sigma_{\xi t}^2$ is known, and zero in the data analysis.
Hence the functions and parameters need to be estimated are given by
$f_0(\cdot)$, $f_1(\cdot)$, $\bm{\theta}_t\equiv(\bm{\alpha}'_t,\bm{\beta}'_t,v^2_t,\lambda_t)'$; $t=1,\dots,T$, and $\bm{a}$.
For $t=1,\dots,T$, let $\bm{\Phi}_t$ be an $n\times K$ matrix with the $(i,k)$-th entry $\varphi_k(\bm{s}_i)$,
and $\bm{X}_t$ be an $n\times p$ matrix with the $(i,\ell)$-th entry $x_{t \ell}(\bm{s}_i)$.
Then from \eqref{eq:data1} and \eqref{eq:model},
the AirBox observations can be rewritten as:
\begin{equation}
\bm{z}_t=\bm{\Phi}_t\bm{\alpha}_t+\bm{X}_t\bm{\beta}_t+\bm{\eta}_t+\bm{\varepsilon}_t;\quad t=1,\dots,T,
\label{eq:data3}
\end{equation}

\noindent where $\bm{\eta}_t\equiv(\eta_t(\bm{s}_1),\dots,\eta_t(\bm{s}_n))'$ and
$\bm{\varepsilon}_t\equiv(\varepsilon_t(\bm{s}_1),\dots,\varepsilon_t(\bm{s}_n))'$.
This, together with \eqref{eq:data2}-\eqref{eq:model},
gives a complete picture of our model.

The model containing many parameters is flexible and can capture spatial heterogeneities.
How to estimate them are illustrated in the following three sub-sections.
How to select $K$ (i.e., the number of basis functions) is discussed in Section~\ref{sec:selection}.
Since there isn't much gain in statistical efficiency to consider a full likelihood approach when data are plenty,
and it is essential to account for outliers,
we estimate the parameters in steps using several robust methods.
For ease of notation, we provide detailed formulae only for fully observed data.

\subsubsection{Estimation of regression parameters $\{\bm{\alpha}_t\}$ and $\{\bm{\beta}_t\}$}
\label{sec:estimation1}

From \eqref{eq:data3},
for each $t=1,\dots,T$, we estimate $\bm{\alpha}_t$ and $\bm{\beta}_t$  using Huber's M-estimator (Huber and Ronchetti, 2009):
\[
\big(\hat{\bm{\alpha}}_t,\hat{\bm{\beta}}_t\big)=\mathop{\arg\min}_{(\bm{\alpha},\bm{\beta})}\sum_{i=1}^n
\rho\big(\delta_t(\bm{s}_i)\big/\sigma_{\delta t}\big);\quad t=1,\dots,T,
\]
where $\delta_t(\bm{s}_i)\equiv z_t(\bm{s}_i)-\bm{\varphi}(\bm{s}_i)'\bm{\alpha}-\bm{x}_t(\bm{s}_i)'\bm{\beta}$,
\[
\sigma_{\delta t}\equiv\mathrm{MAD}\big(\delta_t(\bm{s}_1),\dots,\delta_t(\bm{s}_n)\big)
\equiv\frac{1}{\Psi^{-1}(0.75)}\mathrm{median}\big(\big|\{\delta_t(\bm{s}_i)\}-\mathrm{median}(\{\delta_t(\bm{s}_i)\})\big|\big)
\]
is a robust estimate of the standard deviation of $\delta_t(\bm{s}_i)$ based on the median absolute deviation (MAD),
$\Psi(\cdot)$ is the cumulative distribution function of the standard normal distribution, and
\[
\rho(x)=\left\{
\begin{array}{ll}
\displaystyle\frac{1}{2}x^2; & \mbox{if }|x|\leq c,\\
c|x|-\displaystyle\frac{1}{2}c^2; & \mbox{if }|x|>c,
\end{array}
\right.
\]
is Huber's function. Here $c=1.345$ is commonly chosen, which gives an efficiency of 95\% if the regression errors are normally distributed.
The resulting residuals are given by
\begin{equation}
\hat{\bm{\delta}}_t\equiv\big(\hat{\delta}_t(\bm{s}_1),\dots,\hat{\delta}_t(\bm{s}_n)\big)'=
\bm{z}_t-\bm{\Phi}'_t\hat{\bm{\alpha}}_t-\bm{X}_t\hat{\bm{\beta}}_t;\quad t=1,\dots,T.
\label{eq:delta}
\end{equation}

\noindent In our data analysis, we consider no covariates $\bm{X}_t$ and select $K=25$ basis functions determined by the locations of the AirBox data at time $t$ to obtain $\bm{\Phi}_t$.

\subsubsection{Estimation of spatial covariance parameters $\{v_t^2\}$ and $\{\lambda_t\}$}
\label{sec:estimation2}

Since from \eqref{eq:sigmasq}, $\{\sigma_{\varepsilon t}^2(\bm{s}_i)\}$ are heterogeneous in space,
the conventional variogram approach cannot be applied directly to estimate $v_t$ and $\lambda_t$.
Instead, for each $t=1,\dots,T$, we propose to estimate $v^2_t$ and $\lambda_t$, by fitting the covariances:
\[
\tau_{ht}\equiv v^2_t\exp(-h/\lambda_t);\quad h\in\mathcal{H},
\]
based on a highly robust minimum covariance determinant estimator $\hat{\tau}_{ht}$ of Rousseeuw and van Driessen (1999)
at $h\in\mathcal{H}$ with $\mathcal{H}$ a pre-specified set of distances.
Specifically, for $t=1,\dots,T$ and $h\in\mathcal{H}$, we obtain $\hat{\tau}_{ht}$ based on $\{\big(\hat{\delta}_t(\bm{s}_i),\hat{\delta}_t(\bm{s}_k)\big):(i,k)\in\mathcal{T}_h\}$ with
$\mathcal{T}_h\equiv\{(i,k):h-\Delta<\|\bm{s}_i-\bm{s}_k\|\leq h+\Delta\}$ a tolerance region consisting of pairs distanced between $h+\Delta$ and $h+\Delta$.
After $\{\hat{\tau}_{ht}\}$ are obtained,
we estimate $v^2_t$ and $\lambda_t$ by the constrained least-squares estimators:
\[
\big(\hat{v}^2_t,\hat{\lambda}_t\big)'\equiv\mathop{\arg\min}_{(v^2,\lambda)'\in(0,\infty)^2}\sum_{h\in\mathcal{H}}
\{\hat{\tau}_{ht}-v^2\exp(-h/\lambda)\}^2;\quad t=1,\dots,T.
\]
In our data analysis, we select $\mathcal{H}=\{1/2,2/2,\dots,49/2\}$ (in km) and $\Delta=1/2$ (in km).

\subsubsection{Estimation of measurement-error variance parameters $\bm{a}$}
\label{sec:estimation3}

From \eqref{eq:sigmasq}, we estimate the heterogeneous measurement-error variances
by utilizing a particular subset of the AirBox data in 2020.
This dataset contains 25 AirBoxes at a common location $\bm{s}_0$ (with longitude $121.451^{\circ}$ E and latitude $25.062^{\circ}$ N),
enabling us to focus on measurement-error variances with no other confounding factors.
As before, we aggregate the AirBox data into hourly data.
Because the data are colocated at $\bm{s}_0$,
it is reasonable to assume that the discrepancies between AirBox measurements are fully contributed by measurement errors.

We first check the consistency of measurements from these 25 AirBoxes by computing their mutual sample correlation coefficients.
We remove observations from 13 AirBoxes, two of which have no records and eleven of which have sample correlations with the others all smaller than $0.85$.
For $t=1,\dots,T$, let $\{w_{tj}:j\in\mathcal{J}_t\}$ be the observations available from the remaining 12 AirBoxes,
where $\mathcal{J}_t$ is the corresponding index set.
Then a robust estimate of $\sigma_{\varepsilon t}(\bm{s}_0)$ in \eqref{eq:sigmasq} is given by
$\tilde{\sigma}_{\varepsilon t}\equiv\mathrm{MAD}\big(\big\{w_{tj}:j\in\mathcal{J}_t \big\}\big)$, for $t=1,\dots,T$.

Let $\tilde{z}_{t}$ be the $10\%$ trimmed mean of $\{w_{tj}:j\in\mathcal{J}_t\}$.
We then estimate $\bm{a}\equiv(a_0,a_1,a_2,a_3)'$ in \eqref{eq:sigmasq} by regressing $\{\tilde{\sigma}^2_{\varepsilon t}\}$ on $\{\tilde{z}_{t}\}$ using the following constrained least-squares estimator:
\begin{align}
\hat{\bm{a}}\equiv(\hat{a}_{0},\hat{a}_{1},\hat{a}_{2},\hat{a}_{3})'=\mathop{\arg\min}_{\bm{a}\in[0,\infty)^4}\sum_{t=1}^{T}  
\big\{\tilde{\sigma}_{\varepsilon t }^2-a_0-a_1\tilde{z}_{t}-(a_2-a_1)(\tilde{z}_{t}-a_3)_{+}\big\}^2.
\label{eq:a0a1}
\end{align}

\begin{figure}\centering
\includegraphics[scale=1,trim={0cm 0cm 0cm 0cm},clip]{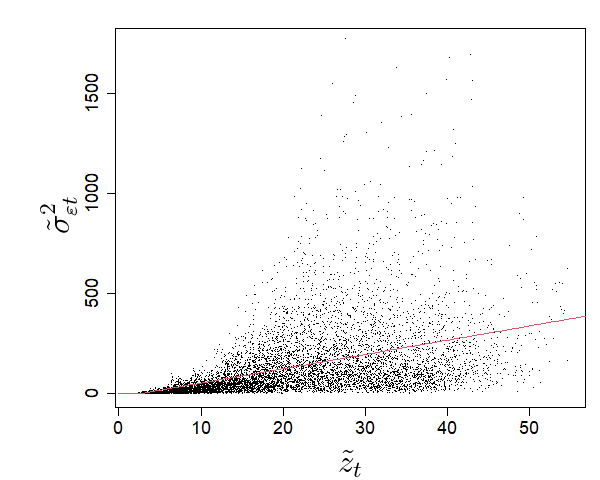}
\caption{The scatter plot of $\{\tilde{\sigma}_{\varepsilon t j}^2\}$ versus $\{\tilde{z}_{tj}\}$ based on the data observed in the year 2020,
where the solid line is the fitted piecewise-linear regression line given in \eqref{eq:a0a1}.}
\label{fig:me}
\end{figure}

\noindent Finally, the proposed estimator of $\sigma^2_{\varepsilon t}(\bm{s}_i)$ is given by
\[
\hat{\sigma}^2_{\varepsilon t}(\bm{s}_i)=\hat{a}_{0}+\hat{a}_{1}z_t(\bm{s}_i)
+(\hat{a}_2-\hat{a}_1)(\tilde{z}_{t}-\hat{a}_3)_{+};\quad i=1,\dots,n,\,t=1,\dots, T.
\]
The estimator is guaranteed to be monotonically non-decreasing in $\tilde{z}_t$ with a slope change at $\hat{a}_3$.
Figure \ref{fig:me} shows the scatter plot of $\{\tilde{\sigma}_{\varepsilon t}^2\}$ versus $\{\tilde{z}_{t}\}$
and the fitted piecewise linear regression line based on the data observed in 2020,
where $\hat{a}_0=0$ and the two slopes are $\hat{a}_1=0.54$ and $\hat{a}_2=7.13$ with the slope change at $\hat{a}_3=2.64$.

\subsubsection{Estimation of $f_0(\cdot)$ and $f_1(\cdot)$}
\label{sec:estimation4}

We develop a three-step procedure to estimate $f_0(\cdot)$ and $f_1(\cdot)$
by first obtaining predictors of $\{y_t(\bm{s}^*_j):j=1,\dots,m,\,t=1,\dots,T\}$,
followed by estimating $f_0(\bm{s}^*_j)$ and $f_1(\bm{s}^*_j)$ for $j=1,\dots,m$,
and then applying a kriging method to estimate $f_0(\cdot)$ and $f_1(\cdot)$.

First, for $t=1,\dots,T$, the empirical best linear predictor of $y_t(\bm{s})$ based on $\bm{z}_t$ with the estimated parameters,
$\hat{\bm{a}}$, $\hat{\bm{\alpha}}_t$, $\hat{\bm{\beta}}_t$,  $\hat{v}^2_t$ and $\hat{\lambda}_t$ plugged-in, is given by
\begin{equation}
\tilde{y}_t(\bm{s})=\bm{\varphi}(\bm{s})'\hat{\bm{\alpha}}_t+\bm{x}(\bm{s})'\hat{\bm{\beta}}_t
+\hat{\bm{c}}_t(\bm{s})\big(\hat{\bm{\Sigma}}_{\eta t}+\hat{\bm{\Sigma}}_{\varepsilon t}\big)^{-1}(\bm{z}_t-\bm{\Phi}_t\hat{\bm{\alpha}}_t-\bm{X}_t\hat{\bm{\beta}}_t),
\label{eq:uk}
\end{equation}

\noindent where
$\hat{\bm{c}}_t(\bm{s})\equiv\big(\hat{v}^2_t
\exp\big(-\|\bm{s}-\bm{s}_1\|/\hat{\lambda}_t\big),\dots,\hat{v}^2_t\exp\big(-\|\bm{s}-\bm{s}_n\|/\hat{\lambda}_t\big)\big)'$
is an estimator of 
$\bm{c}_t(\bm{s})\equiv\big(v^2_t\exp\big(-\|\bm{s}-\bm{s}_1\|/\lambda_t\big),\dots,v^2_t\exp\big(-\|\bm{s}-\bm{s}_n\|/\lambda_t\big)\big)'$,
$\hat{\bm{\Sigma}}_{\eta t}$ is an $n\times n$ matrix with the $(i,j)$-th entry $\hat{v}^2_t\exp\big(-\|\bm{s}_i-\bm{s}_j\|/\hat{\lambda}_t\big)$, and
$\hat{\bm{\Sigma}}_{\varepsilon t}\equiv\mathrm{diag}\big(\hat{\sigma}_{\varepsilon t}^2(\bm{s}_1),\dots,\hat{\sigma}_{\varepsilon t}^2(\bm{s}_n)\big)$.

Next, we estimate $f_0(\bm{s}^*_j)$ and $f_1(\bm{s}^*_j)$ in \eqref{eq:calibration}, for $j=1,\dots,m$,
by applying a regression calibration method.
If it is reasonable to assume that both $f_0(\cdot)$ and $f_1(\cdot)$ are constant functions,
then it suffices to consider a global calibration line by simply regressing $\{z^*_t(\bm{s}^*_j):j=1,\dots,m,\,t=1,\dots,T\}$ on $\big\{\tilde{y}_t(\bm{s}^*_j):j=1,\dots,m,\,t=1,\dots,T\big\}$,
leading to the ordinary-least-squares (OLS) estimators for the two constant functions of $f_0(\cdot)$ and $f_1(\cdot)$:
\begin{equation}
  \tilde{f}^{(g)}_0
= \bar{z}-\tilde{f}^{(g)}_1\bar{y},\quad
  \tilde{f}^{(g)}_1
= \frac{\sum_{t=1}^T\sum_{j=1}^m\big(\tilde{y}_t(\bm{s}^*_j)-\bar{y}\big)\big(z^*_t(\bm{s}^*_j)-\bar{z}\big)}{\sum_{t=1}^T\sum_{j=1}^m\big(\tilde{y}_t(\bm{s}^*_j)-\bar{y}\big)^2},
\label{eq:OLS}
\end{equation}

\noindent where $\bar{y}\equiv\sum_{t=1}^T\sum_{j=1}^m \tilde{y}_t(\bm{s}^*_j)/(mT)$ and $\bar{z}\equiv\sum_{t=1}^T\sum_{j=1}^m z^*_t(\bm{s}^*_j)/(mT)$.

To achieve spatially adaptive calibration,
we first regress $\{z^*_t(\bm{s}^*_j):t=1,\dots,T\}$ on $\big\{\tilde{y}_t(\bm{s}^*_j):t=1,\dots,T\big\}$,
and obtain preliminary estimators $\tilde{f}_0(\bm{s}^*_j)$ and $\tilde{f}_1(\bm{s}^*_j)$ of $f_0(\bm{s}^*_j)$ and $f_1(\bm{s}^*_j)$
at the data locations $\{\bm{s}^*_j:j=1,\dots,m\}$ using OLS.
Let $v_0(\bm{s}^*_j)$ and $v_1(\bm{s}^*_j)$ be the estimated standard errors of $\tilde{f}_0(\bm{s}^*_j)$ and $\tilde{f}_1(\bm{s}^*_j)$, respectively.
We then treat $f_0(\cdot)$ as a spatial process and consider the ordinary-kriging predictor $\hat{f}_0(\bm{s})$ of $f_0(\bm{s})$, for $\bm{s}\in D$.
Specifically, the ordinary-kriging predictor is obtained using the isotropic exponential covariance model (estimated by maximum likelihood)
based on the following measurement equation:
\[
\tilde{f}_0(\bm{s}^*_j)=f_0(\bm{s}^*_j)+u_0(\bm{s}^*_j);\quad j=1,\dots,m,
\]
where $u_0(\bm{s}^*_j)\sim N(0,v^2_0(\bm{s}^*_j))$; $j=1,\dots,m$, are independent noise variables.
Similar treatment is applied to $f_1(\bm{s})$,
and obtain the ordinary-kriging predictor $\hat{f}_1(\bm{s})$ of $f_1(\bm{s})$, for $\bm{s}\in D$.

\subsection{Spatial prediction based on calibrated AirBox data}

Utilizing the model given by \eqref{eq:data1}, \eqref{eq:calibration} and \eqref{eq:model},
the best linear predictor of $y^*_t(\cdot)$
given $f_0(\cdot)$, $f_1(\cdot)$, $\bm{\theta}_t$ and $\bm{a}$ is
    \begin{align}
      \tilde{y}^*_t(\bm{s}; f_0,f_1, & \bm{\theta}_t,\bm{a})
    \equiv \mathrm{E}(y^*_t(\bm{s})|\bm{z}_t)\notag\\
    =&~ f_0(\bm{s})+f_1(\bm{s})\big(\bm{\varphi}(\bm{s})'\bm{\alpha}_t+\bm{x}_t(\bm{s})'\bm{\beta}_t\big)\notag\\
    &~ +f_1(\bm{s})\bm{c}_t(\bm{s})\big(\bm{\Sigma}_{\eta t}+\bm{\Sigma}_{\varepsilon t}\big)^{-1}(\bm{z}_t-\bm{\Phi}_t\bm{\alpha}_t-\bm{X}_t\bm{\beta}_t);\quad
      \bm{s}\in D,\,t=1,\dots,T,
      \label{eq:ok2}
    \end{align}
where $\bm{\Sigma}_{\eta t}$ is an $n\times n$ matrix with the $(i,j)$-th entry $v^2_t\exp\big(-\|\bm{s}_i-\bm{s}_i\|/\lambda_t\big)$, and
$\bm{\Sigma}_{\varepsilon t}\equiv\mathrm{diag}\big(\sigma_{\varepsilon t}^2(\bm{s}_1),\dots,\sigma_{\varepsilon t}^2(\bm{s}_n)\big)$.
The corresponding mean-squared prediction error (i.e., kriging variance) for $\bm{s}\in D$ and $t=1,\dots,T$ is
    \begin{equation}
    \mathrm{E}\big(\tilde{y}^*_t(\bm{s};f_0,f_1,\bm{\theta}_t,\bm{a})-y_t(\bm{s})\big)^2=f_1(\bm{s})^2\big\{v^2_t-\bm{c}_t(\bm{s})'
    \big(\bm{\Sigma}_{\eta t}+\bm{\Sigma}_{\varepsilon t}\big)^{-1}\bm{c}_t(\bm{s})\big\}.
    \label{eq:okvar2}
    \end{equation}

\noindent After plugging-in the estimated $\hat{f}_0(\cdot)$, $\hat{f}_1(\cdot)$, $\hat{\bm{\theta}}_t$, and $\hat{\bm{a}}$ in \eqref{eq:ok2},
the proposed predictor of $y^*_t(\bm{s})$ for $\bm{s}\in D$ based only on AirBox data is given by
\begin{equation}
\tilde{y}^*_t(\bm{s};\hat{f}_0,\hat{f}_1,\hat{\bm{\theta}}_t,\hat{\bm{a}});\quad \bm{s}\in D,\,t=1,\dots,T,
\label{eq:ok3}
\end{equation}
where $\hat{\bm{\theta}}_t\equiv(\hat{\bm{\alpha}}'_t,\hat{\bm{\beta}}'_t,\hat{v}^2_t,\hat{\lambda}_t)'$; $t=1,\dots,T$.

\subsection{Selection of $K$}
\label{sec:selection}

As demonstrated in Huang \textit{et al}.~(2018), the spatial prediction is not much affected by $K$,
since both the basis functions $\bm{\varphi}(\cdot)$ and the spatial process $\eta_t(\cdot)$ compete to capture $y_t(\cdot)$.
Similar to Huang \textit{et al}.~(2018), we select $K=25$ so that the function in the projected space accounts for about 50\% of the variation in the AirBox data.
Alternatively, $K$ can be selected by using the conditional Akaike's information criterion of Vaida and Blanchard (2005)
or cross validation.

\subsection{Spatial prediction combining EPA and AirBox data}
\label{sec:combine}

For the model given by \eqref{eq:data1}-\eqref{eq:model},
the best linear predictor of $y^*_t(\bm{s})$ incorporating both $\bm{z}_t$ and $\bm{z}^*_t$
with given $f_0(\cdot)$, $f_1(\cdot)$, $\bm{\theta}_t$ and $\bm{a}$ is
    \begin{align}
      \hat{y}^*_t(\bm{s};
    \,&  f_0,f_1,\bm{\theta}_t,\bm{a})\equiv\mathrm{E}(y^*_t(\bm{s})|\bm{z}_t,\bm{z}^*_t)\notag\\
    =&~ f_0(\bm{s})+f_1(\bm{s})\big(\bm{\varphi}_t(\bm{s})'\bm{\alpha}_t+\bm{x}_t(\bm{s})'\bm{\beta}_t\big)\notag\\
  &~ +
      f_1(\bm{s})\bm{c}^*_t(\bm{s})'
      \left( 
      \left(
     \begin{matrix}
      \bm{F}_1 & \bm{0}\\
      \bm{0} & \bm{I}
      \end{matrix}
      \right)  \bm{\Sigma}^*_{\eta t}      \left(
     \begin{matrix}
      \bm{F}_1 & \bm{0}\\
      \bm{0} & \bm{I}
      \end{matrix}
      \right) 
     +\left(
      \begin{matrix}
      \sigma^2_{\xi t}\bm{I} & \bm{0}\\
      \bm{0} & \bm{\Sigma}_{\varepsilon t}
      \end{matrix}
      \right)\right)^{-1}\notag\\
    &~ \times\left(
      \begin{matrix}
      \bm{z}^*_t-\bm{f}_0-\bm{F}_1(\bm{\varphi}(\bm{s}^*_1),\dots,\bm{\varphi}(\bm{s}^*_m))'\bm{\alpha}_t-
      \bm{F}_1(\bm{x}_t(\bm{s}^*_1),\dots,\bm{x}_t(\bm{s}^*_m))'\bm{\beta}_t\\
      \bm{z}_t-(\bm{\varphi}(\bm{s}_{1}),\dots,\bm{\varphi}(\bm{s}_{n}))'\bm{\alpha}_t-
      (\bm{x}_t(\bm{s}_{1}),\dots,\bm{x}_t(\bm{s}_{n}))'\bm{\beta}_t
      \end{matrix}
      \right),
      \label{eq:calibrate}
    \end{align}

\noindent for $\bm{s}\in D$ and $t=1,\dots,T$, where
    \begin{align*}
      \bm{c}^*_t(\bm{s})
    \equiv&~\big(f_1(\bm{s}^*_1)v^2_t\exp\big(-\|\bm{s}-\bm{s}^*_1\|/\lambda_t\big),\dots,f_1(\bm{s}^*_m)v^2_t\exp\big(-\|\bm{s}-\bm{s}^*_m\|/\lambda_t\big),\bm{c}'_t(\bm{s})\big)',\\
      \bm{\Sigma}^*_{\eta t}
    \equiv&~ \mathrm{var}\big(\eta_t(\bm{s}^*_1),\dots,\eta_t(\bm{s}^*_m),\eta_t(\bm{s}_1),\dots,\eta_t(\bm{s}_n)\big),
    \end{align*}

\noindent $\bm{f}_0\equiv(f_0(\bm{s}^*_1),\dots,f_0(\bm{s}^*_m))'$ and $\bm{F}_1\equiv\mathrm{diag}(f_1(\bm{s}^*_1),\dots,f_1(\bm{s}^*_m))$.
The corresponding mean squared prediction error (i.e., kriging variance) is
    \begin{align}
    \mathrm{E}\big(\hat{y}^*_t(\bm{s};f_0,
    & f_1,\bm{\theta}_t,\bm{a})-y^*_t(\bm{s})\big)^2\notag\\
    =&~ f_1(\bm{s})^2\left\{v^2_t-\bm{c}^*_t(\bm{s})'
    \left(\left(\begin{matrix}
    \bm{F}_1 & \bm{0}\\
    \bm{0} & \bm{I}
    \end{matrix}
    \right)\bm{\Sigma}_{\eta t}^*\left(
    \begin{matrix}
    \bm{F}_1 & \bm{0}\\
    \bm{0} & \bm{I}
    \end{matrix}
   \right)+\left(
    \begin{matrix}
    \sigma^2_{\xi t}\bm{I} & \bm{0}\\
    \bm{0} & \bm{\Sigma}_{\varepsilon t}
    \end{matrix}
    \right)\right)^{-1}
    \bm{c}^*_t(\bm{s})\right\},
    \label{eq:calibrate.var}
    \end{align}
for $\bm{s}\in D$ and $t=1,\dots,T$.

\subsection{Diagnostics}

We conduct model diagnostics using the standardized residuals for $t=1,\dots,T$:
\begin{align}
  r_t(\bm{s}_i)
\equiv&~ \frac{z_t(\bm{s}_i)-\hat{y}_t\big(\bm{s}_i;\hat{f}_0,\hat{f}_1,\hat{\bm{\theta}}_t,\hat{\bm{a}}\big)}
  {\sigma_t\big(\bm{s}_i;\hat{f}_0,\hat{f}_1,\hat{\bm{\theta}_t},\hat{\bm{a}}\big)};\quad i=1,\dots,n,
\label{eq:res1}
\end{align}

\noindent where $\hat{y}_t(\bm{s};f_0,f_1,\bm{\theta}_t,\bm{a})\equiv f_1(\bm{s})^{-1}\big\{-f_0(\bm{s})+\hat{y}^*_t(\bm{s};f_0,f_1,\bm{\theta}_t,\bm{a})\big\}$; $\bm{s}\in D$,
\begin{align*}
  \sigma^2_t(\bm{s}_i;f_0,f_1,
& \bm{\theta}_t,\bm{a})
\equiv \mathrm{var}(z_t(\bm{s}_i)-\hat{y}_t(\bm{s}_i;f_0,f_1,\bm{\theta}_t,\bm{a}))\notag\\
=&~ (\bm{0},\bm{e}'_i\bm{\Sigma}_{\varepsilon t})
   \left(\left(\begin{matrix}
    \bm{F}_1 & \bm{0}\\
    \bm{0} & \bm{I}
    \end{matrix}
    \right)\bm{\Sigma}_{\eta t}^*\left(
    \begin{matrix}
    \bm{F}_1 & \bm{0}\\
    \bm{0} & \bm{I}
    \end{matrix}
   \right)+\left(
    \begin{matrix}
    \sigma^2_{\xi t}\bm{I} & \bm{0}\\
    \bm{0} & \bm{\Sigma}_{\varepsilon t}
    \end{matrix}
    \right)\right)^{-1}
      \left(
      \begin{matrix}
      \bm{0}\\ \bm{\Sigma}_{\varepsilon t}\bm{e}_i
      \end{matrix}
      \right),
\end{align*}

\noindent and $\bm{e}_i$ is the $i$-th column of $\bm{I}_n$.
Note that for $t=1,\dots,T$, we have $r_t(\bm{s}_i)\sim N(0,1)$
if $\hat{f}_0(\cdot)$, $\hat{f}_1(\cdot)$, $\hat{\bm{\theta}}_t$ and $\hat{\bm{a}}$ in \eqref{eq:res1} are replaced by their corresponding true parameters.

\section{Data analysis results}

\begin{figure}[tb]\centering
\begin{tabular}{cccccc}
\!\!\!January & \!\!\!Februry & \!\!\!March & \!\!\!April & \!\!\!May & \!\!\!June\\
\!\!\!\!\!\includegraphics[scale=0.16,trim={0cm 0cm 0cm 0cm},clip]{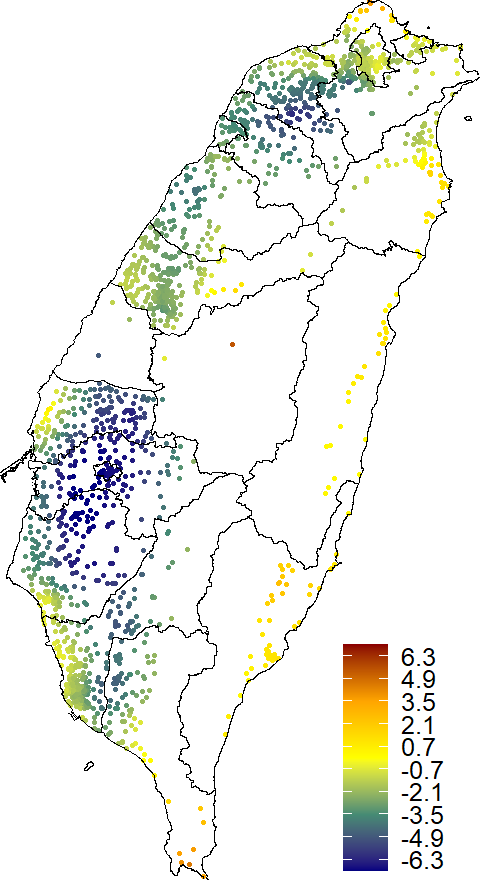}
& \!\!\!\!\!\includegraphics[scale=0.16,trim={0cm 0cm 0cm 0cm},clip]{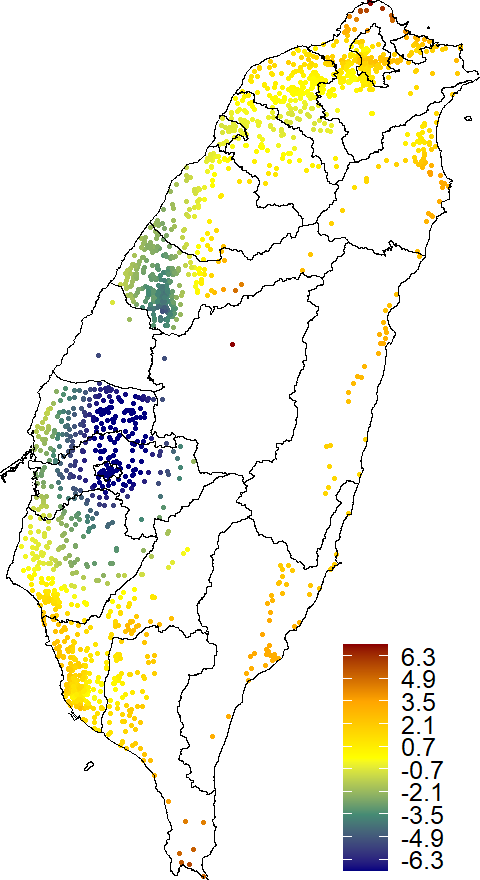}
& \!\!\!\!\!\includegraphics[scale=0.16,trim={0cm 0cm 0cm 0cm},clip]{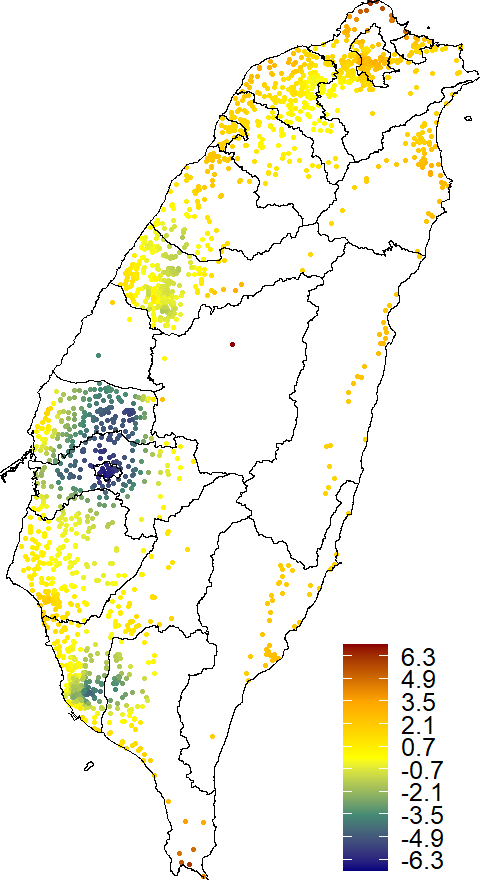}
& \!\!\!\!\!\includegraphics[scale=0.16,trim={0cm 0cm 0cm 0cm},clip]{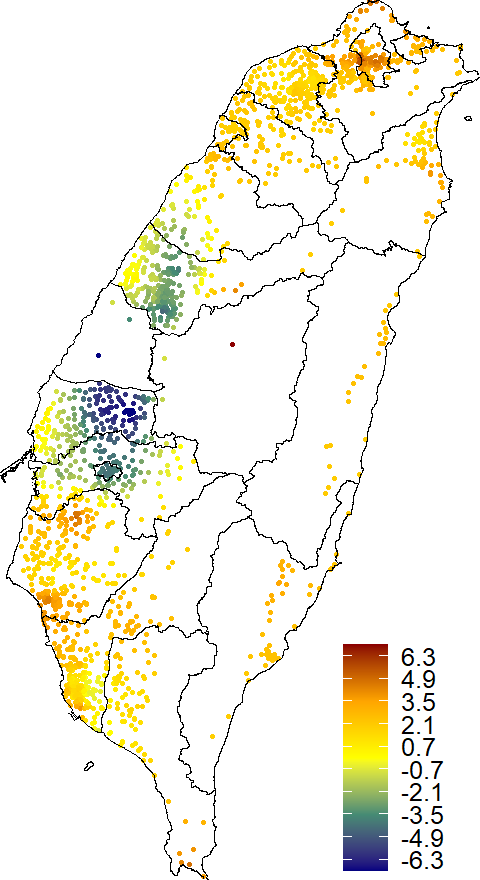}
& \!\!\!\!\!\includegraphics[scale=0.16,trim={0cm 0cm 0cm 0cm},clip]{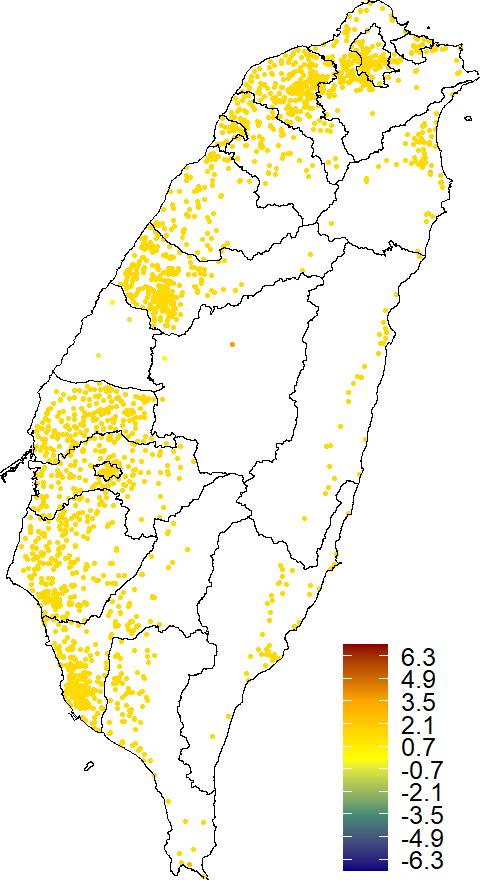}
& \!\!\!\!\!\includegraphics[scale=0.16,trim={0cm 0cm 0cm 0cm},clip]{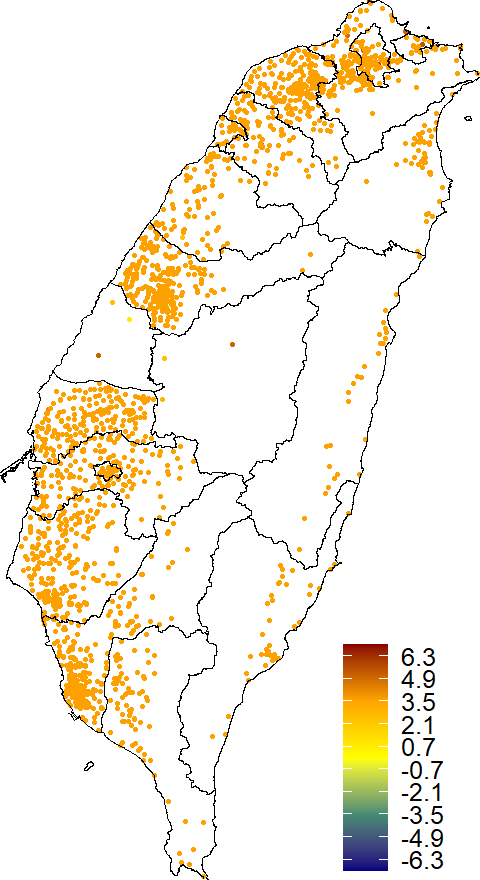}\\
\!\!\!July & \!\!\!August & \!\!\!September & \!\!\!October & \!\!\!November & \!\!\!December\\
\!\!\!\!\!\includegraphics[scale=0.16,trim={0cm 0cm 0cm 0cm},clip]{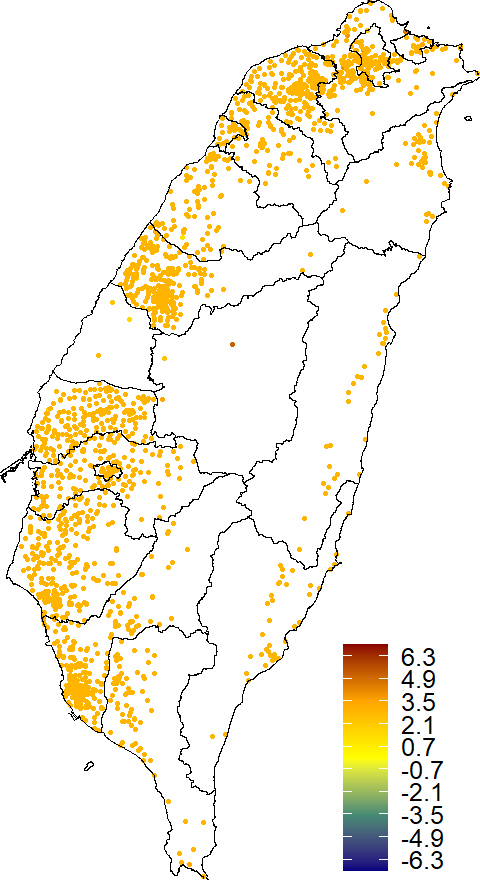}
& \!\!\!\!\!\includegraphics[scale=0.16,trim={0cm 0cm 0cm 0cm},clip]{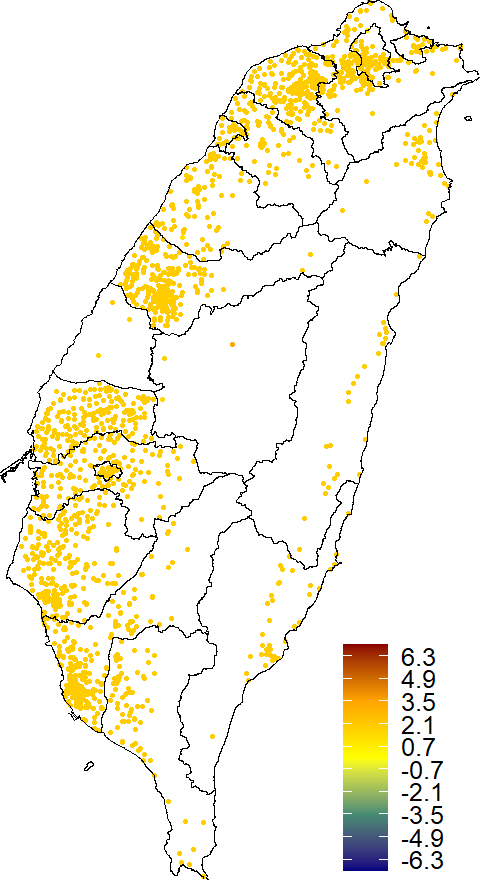}
& \!\!\!\!\!\includegraphics[scale=0.16,trim={0cm 0cm 0cm 0cm},clip]{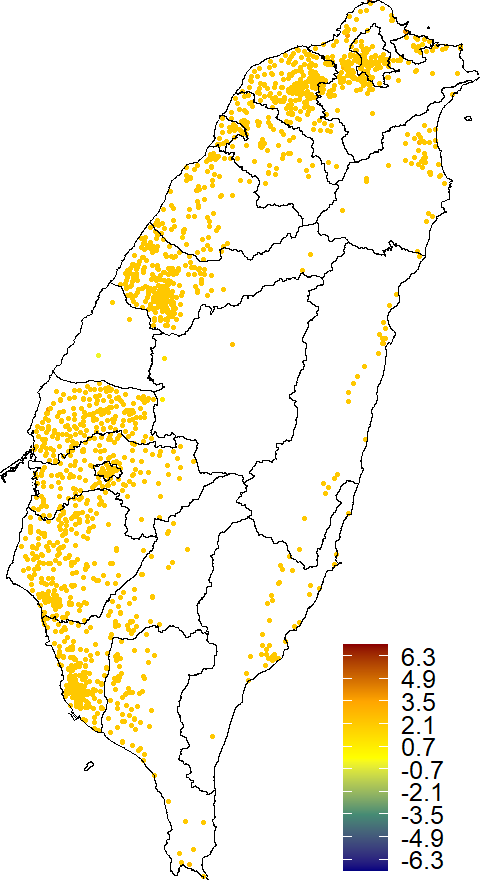}
& \!\!\!\!\!\includegraphics[scale=0.16,trim={0cm 0cm 0cm 0cm},clip]{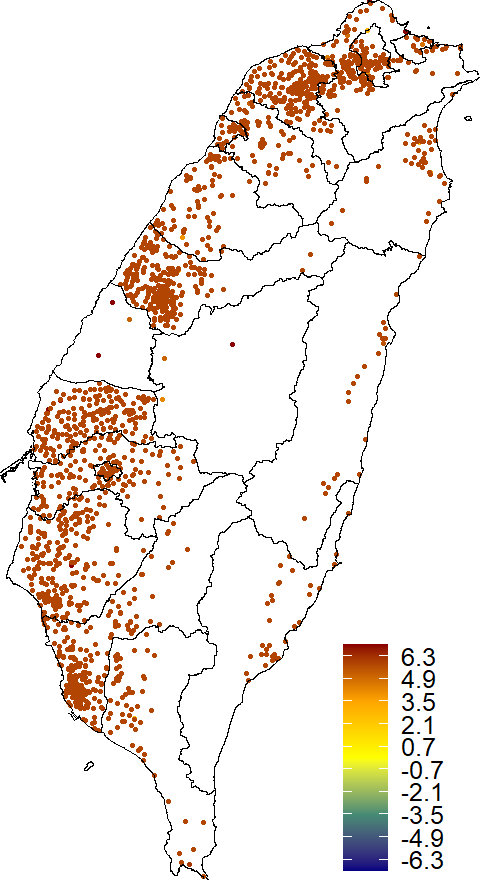}
& \!\!\!\!\!\includegraphics[scale=0.16,trim={0cm 0cm 0cm 0cm},clip]{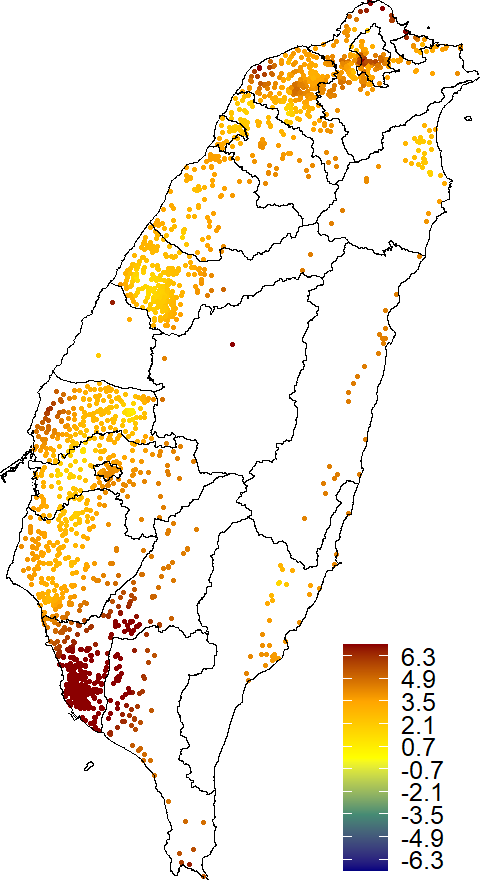}
& \!\!\!\!\!\includegraphics[scale=0.16,trim={0cm 0cm 0cm 0cm},clip]{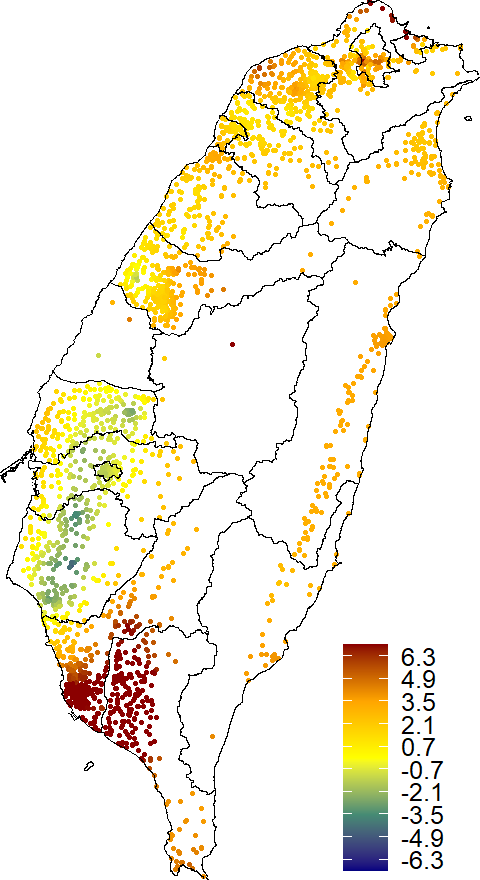}
\end{tabular}
\caption{Month-wise calibrated intercepts for 2020.}
\label{fig:intercept}
\end{figure}

\begin{figure}[tbh]\centering
\begin{tabular}{cccccc}
\!\!\!January & \!\!\!Februry & \!\!\!March & \!\!\!April & \!\!\!May & \!\!\!June\\
\!\!\!\!\!\includegraphics[scale=0.16,trim={0cm 0cm 0cm 0cm},clip]{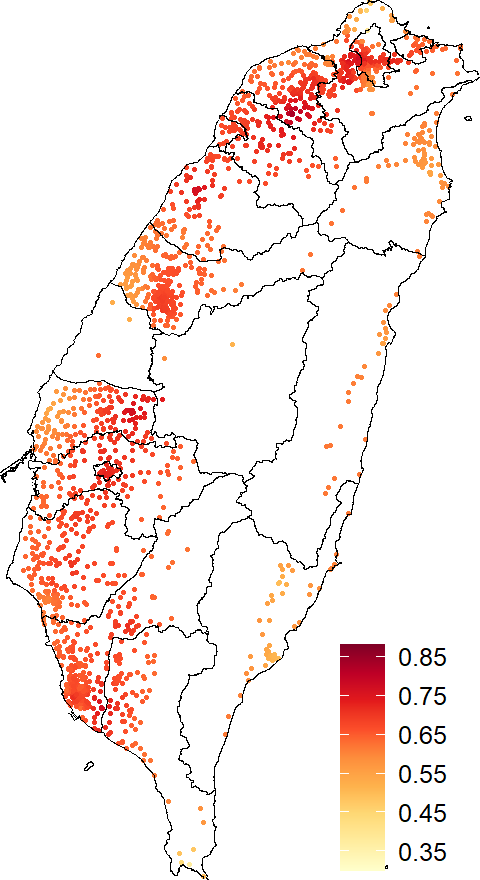}
& \!\!\!\!\!\includegraphics[scale=0.16,trim={0cm 0cm 0cm 0cm},clip]{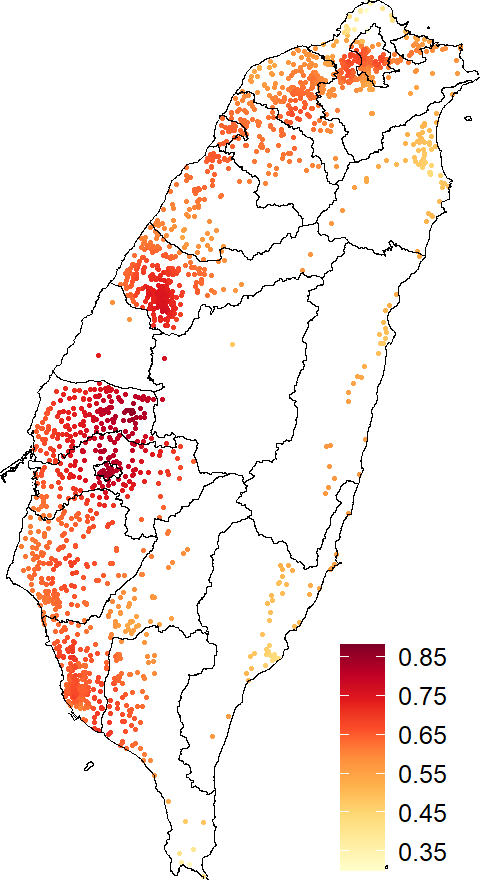}
& \!\!\!\!\!\includegraphics[scale=0.16,trim={0cm 0cm 0cm 0cm},clip]{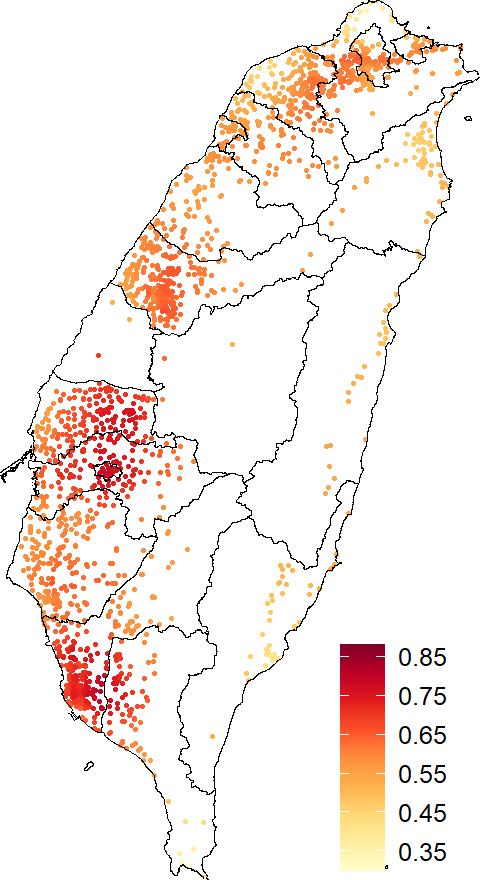}
& \!\!\!\!\!\includegraphics[scale=0.16,trim={0cm 0cm 0cm 0cm},clip]{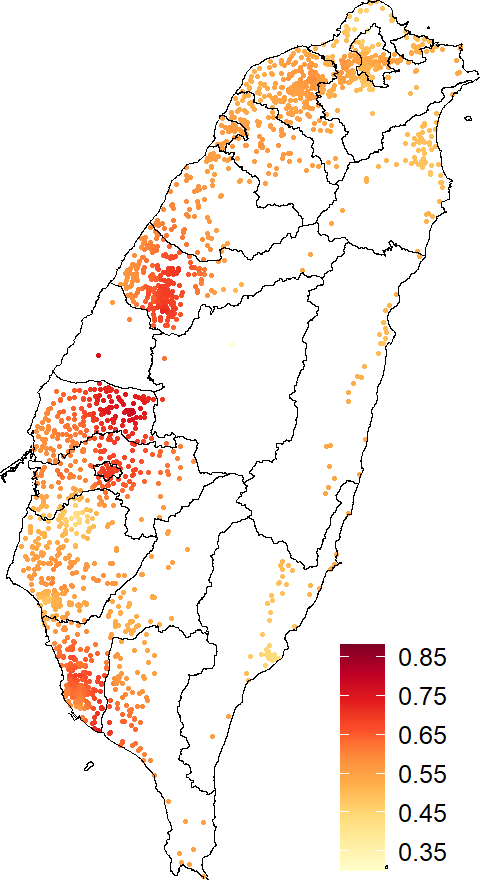}
& \!\!\!\!\!\includegraphics[scale=0.16,trim={0cm 0cm 0cm 0cm},clip]{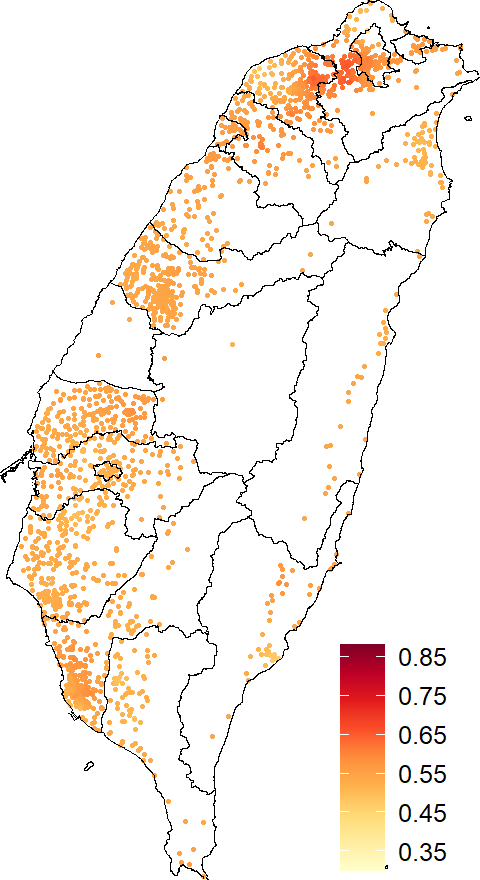}
& \!\!\!\!\!\includegraphics[scale=0.16,trim={0cm 0cm 0cm 0cm},clip]{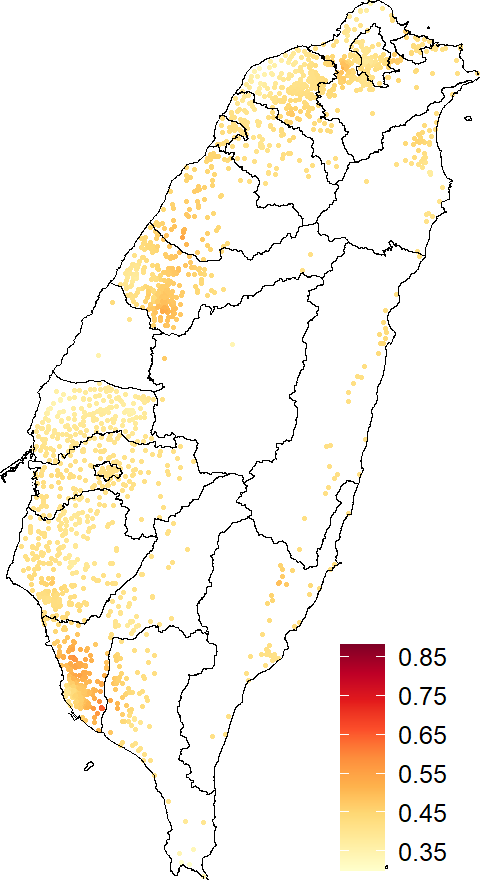}\\
\!\!\!July & \!\!\!August & \!\!\!September & \!\!\!October & \!\!\!November & \!\!\!December\\
\!\!\!\!\!\includegraphics[scale=0.16,trim={0cm 0cm 0cm 0cm},clip]{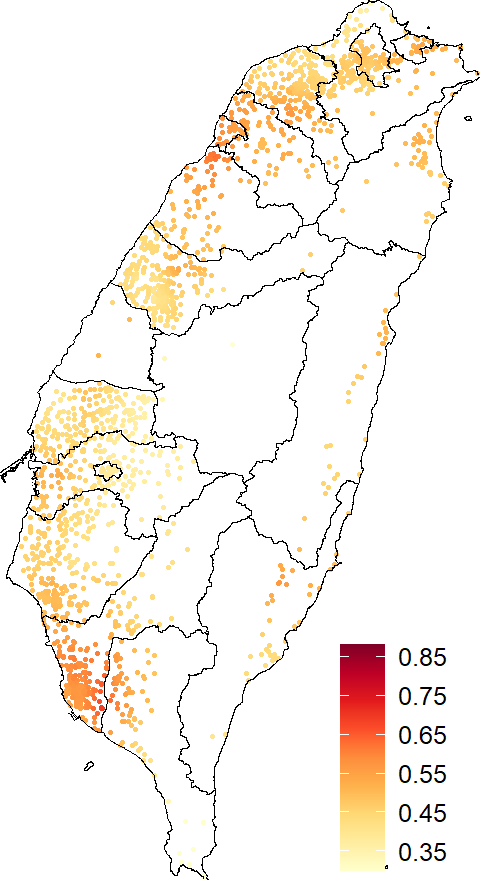}
& \!\!\!\!\!\includegraphics[scale=0.16,trim={0cm 0cm 0cm 0cm},clip]{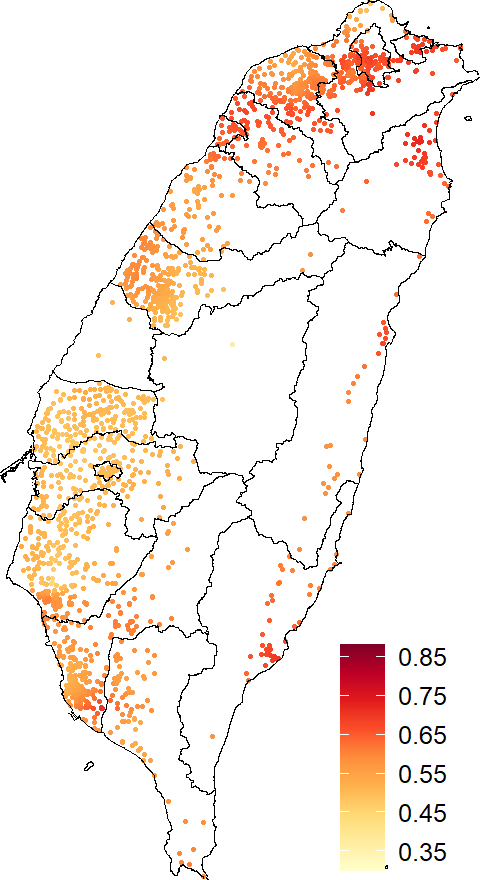}
& \!\!\!\!\!\includegraphics[scale=0.16,trim={0cm 0cm 0cm 0cm},clip]{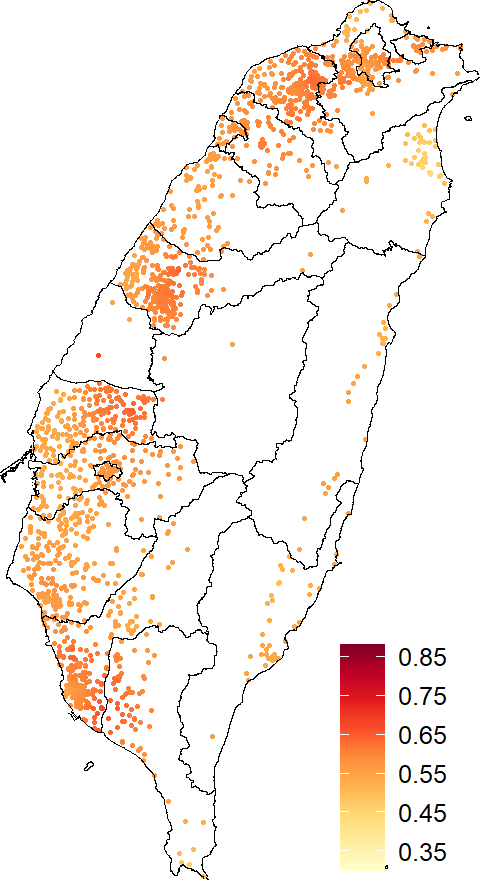}
& \!\!\!\!\!\includegraphics[scale=0.16,trim={0cm 0cm 0cm 0cm},clip]{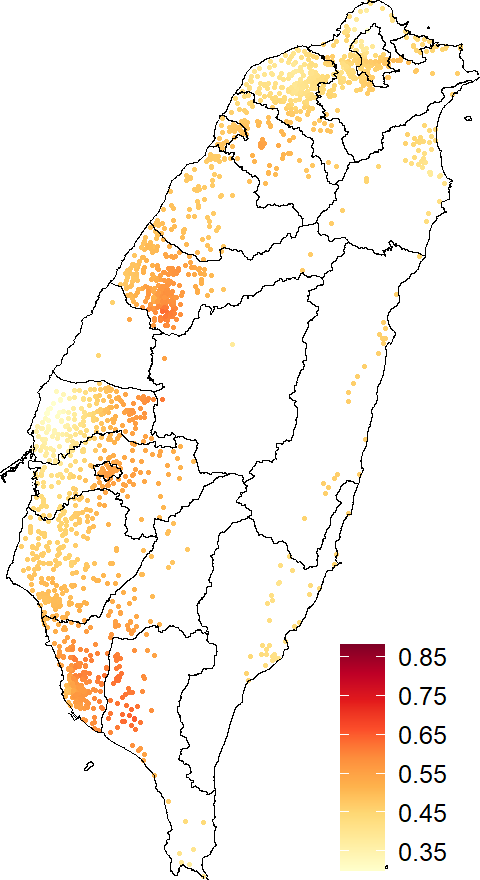}
& \!\!\!\!\!\includegraphics[scale=0.16,trim={0cm 0cm 0cm 0cm},clip]{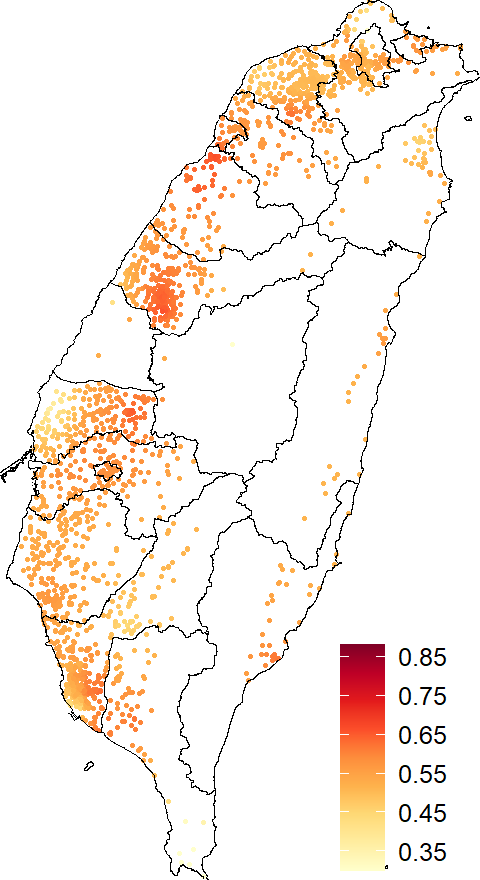}
& \!\!\!\!\!\includegraphics[scale=0.16,trim={0cm 0cm 0cm 0cm},clip]{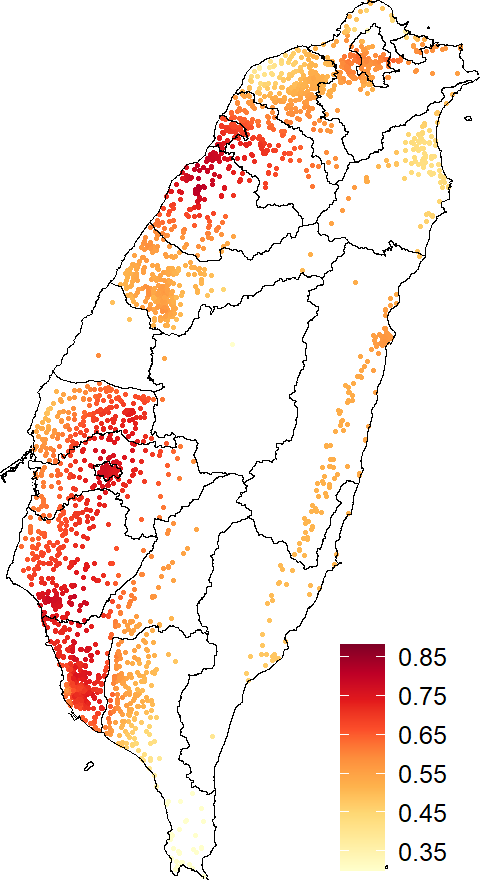}
\end{tabular}
\caption{Month-wise calibrated slopes for 2020.}
\label{fig:slope}
\end{figure}

We applied our method developed in Sections \ref{sec:model}-\ref{sec:combine} to EPA and AirBox data for the year 2020.
There are about respectively  22.2\% and 19.4\% missing observations in the EPA and the AirBox data.
We first did some data cleaning by removing a small portion of unusual PM$_{2.5}$ data that are either negative or larger than 1000 (in ppm) in both datasets.
Among $366\times 24=8784$ hours in the year 2020,
we identified $T=6709$ hours with non-missing observations in at least 500 AirBox locations and 50 EPA locations for model fitting and calibration.
We computed $\hat{\bm{\beta}}_t$,  $\hat{v}_t$ and $\hat{\lambda}_t$ at each hour $t=1,\dots,T$ in 2020 using the proposed method in
Sections \ref{sec:estimation1} and \ref{sec:estimation2} with no covariates $\bm{X}_t$.
We estimated $\bm{a}$ using the method introduced in Section \ref{sec:estimation3}.
Then we estimated $f_0(\cdot)$ and $f_1(\cdot)$ for each month separately using the method described in Section \ref{sec:estimation4}.
The month-wise calibrated functions of intercept and slope are shown in Figures \ref{fig:intercept} and \ref{fig:slope}, respectively.
We can see that both functions vary smoothly in space and time due to
changes in chemical compositions of PM$_{2.5}$ and some other factors, such as
seasonal climate patterns.
The estimated slopes are all less than one because we target measurements from EPA stations located around 10 meters in height, which
tend to produce lower PM$_{2.5}$ values than AirBoxes.
The whole calibrate procedure is computationally fast;
it took less than one hour to obtain $\hat{f}_0(\cdot)$ and $\hat{f}_1(\cdot)$ for each month
(on a PC with AMD Ryzen Threadripper 2920X 12-Core Processor and 64 GB RAM).

\begin{figure}[tbh]\centering
\begin{tabular}{cc}
  \includegraphics[scale=0.32,trim={0cm 0cm 0cm 0cm},clip]{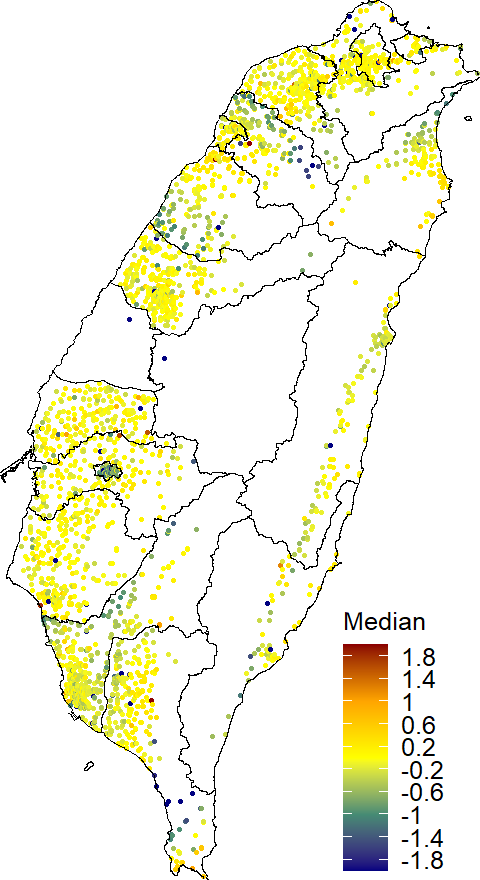}
&  \includegraphics[scale=0.32,trim={0cm 0cm 0cm 0cm},clip]{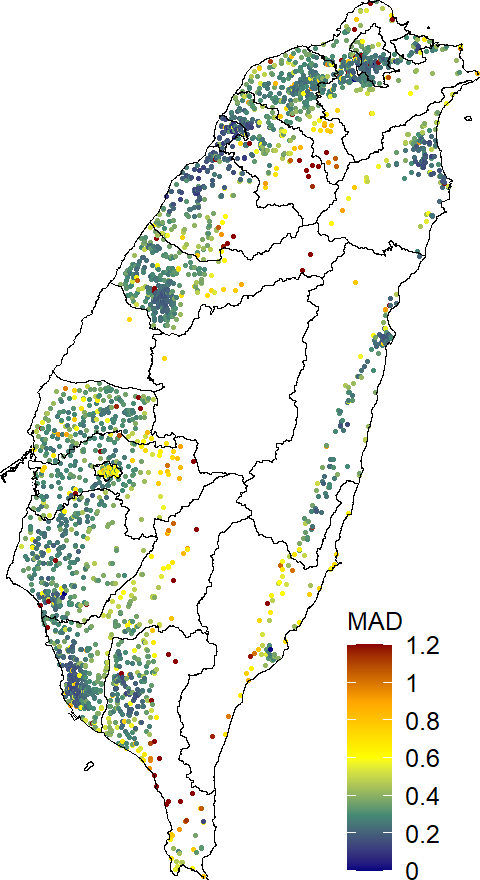}\\
(a) & (b)
\end{tabular}
\caption{(a) Location-wise medians of the standardized residuals; (b) Location-wise MADs of the standardized residuals.}
\label{fig:diagnostic}
\end{figure}

We did a diagnostic check using the standardized residuals of \eqref{eq:res1}.
Ideally, if the model fits the data perfectly, the standardized residuals are approximately standard-normal distributed.
However, it is almost impossible to model the AirBox data perfectly due to many outliers (e.g., some people put their AirBoxes indoors or close to some emission sources,
which are likely to produce unusual small or large measurements, respectively).
Figures \ref{fig:diagnostic}(a) and (b) show the median and the MAD of the standardized residuals at each AirBox location.
The median values are all around zero, showing that the proposed method exhibits a small bias in prediction.
However, the MADs vary in space and are mostly smaller than one, indicating that our method tends to be more conservative in producing prediction intervals.

We also examined the prediction performance among various methods for each month, indexed by $\mathcal{M}_1,\dots,\mathcal{M}_{12}\subset\{1,\dots,T\}$, separately.
Specifically, for each hour $t\in\mathcal{M}_j$, we randomly divided the EPA data (on the main island with 74 stations)
into the training data $\bm{z}^*_t$ (consisting of 2/3 of non-missing observations with various sizes)
and set the remaining data $\bm{z}^{**}_t\equiv\big\{z^{**}_t(\bm{s}^{**}_{t1}),\dots,z^{**}_{t}(\bm{s}^{**}_{t m_t})\big\}$ with $m_t$ locations for testing purpose.
We estimated $\bm{\beta}_t$, $v_t$ and $\lambda_t$ by $\hat{\bm{\beta}}_t$, $\hat{v}_t$ and $\hat{\lambda}_t$ based on $\bm{z}_t$ as before.
But for $j=1,\dots,12$, we computed the estimators of $\bm{a}$, $f_0(\cdot)$ and $f_1(\cdot)$ based only on $\{\bm{z}^{**}_t:t\in\mathcal{M}_j\}$ and $\{\bm{z}_t:t\in\mathcal{M}_j\}$.
We evaluated the prediction performance using the root mean-squared prediction error (RMSPE) criterion based on the test data $\{\bm{z}^{**}_t\}$:
\[
\mathrm{RMSPE}(\mathcal{M}_j)\equiv\bigg\{
\frac{1}{\sum_{t\in\mathcal{M}_j}m_t}\sum_{t\in\mathcal{M}_j}\sum_{k=1}^{m_t}\big|\tilde{\tilde{y}}^*_t(\bm{s}_{tk})-z^{**}_t(\bm{s}_{tk})\big|^2\bigg\}^{1/2};\quad j=1,\dots,12,
\]
where $\tilde{\tilde{y}}^*_t(\bm{s}_{tk})$ is a generic predictor of $y^*_t(\bm{s}_{tk})$; $k=1,\dots,m_t$, $t=1,\dots,T$.

We compared among six methods:
\begin{enumerate}
\item[M1] Apply the predictor \eqref{eq:uk} based on AirBox data alone with no calibration.
\item[M2] Apply the predictor \eqref{eq:ok3} based on AirBox data alone with a monthly global calibration procedure.
  Specifically, $\hat{f}_0(\cdot)$ and $\hat{f}_1(\cdot)$ in \eqref{eq:ok3} are replaced by $\tilde{f}^{(g)}_0(\cdot)$ and $\tilde{f}^{(g)}_1(\cdot)$ in \eqref{eq:OLS}.
\item[M3] Apply the predictor \eqref{eq:calibrate} combining EPA and AirBox data with a monthly global calibration procedure.
  Specifically, $f_0(\cdot)$ and $f_1(\cdot)$ in \eqref{eq:calibrate} are replaced by $\tilde{f}^{(g)}_0(\cdot)$ and $\tilde{f}^{(g)}_1(\cdot)$ in \eqref{eq:OLS}, and
  $\bm{\theta}_t$'s and $\bm{a}$ in \eqref{eq:calibrate} are replaced by $\hat{\bm{\theta}}_t$'s and $\hat{\bm{a}}$.
\item[M4] Apply the predictor \eqref{eq:ok3} based on AirBox data alone with the proposed spatially adaptive calibration procedure for each month.
 \item[M5] Apply the predictor \eqref{eq:calibrate} combining EPA and AirBox data with the proposed spatially adaptive calibration procedure
   for each month,
  where $f_0(\cdot)$, $f_1(\cdot)$, $\bm{\theta}_t$'s and $\bm{a}$ in \eqref{eq:calibrate} are replaced by
  $\hat{f}_0(\cdot)$, $\hat{f}_1(\cdot)$, $\hat{\bm{\theta}}_t$'s and $\hat{\bm{a}}$, respectively.
\item[M6] Apply ordinary kriging by using the isotropic exponential covariance model (including the nugget effect) based on EPA data only for each hour,
  where the model parameters are estimated by maximum likelihood.
\end{enumerate}

\begin{figure}[bt]\centering
\!\!\includegraphics[scale=0.245,trim={0cm 0cm 0cm 0cm},clip]{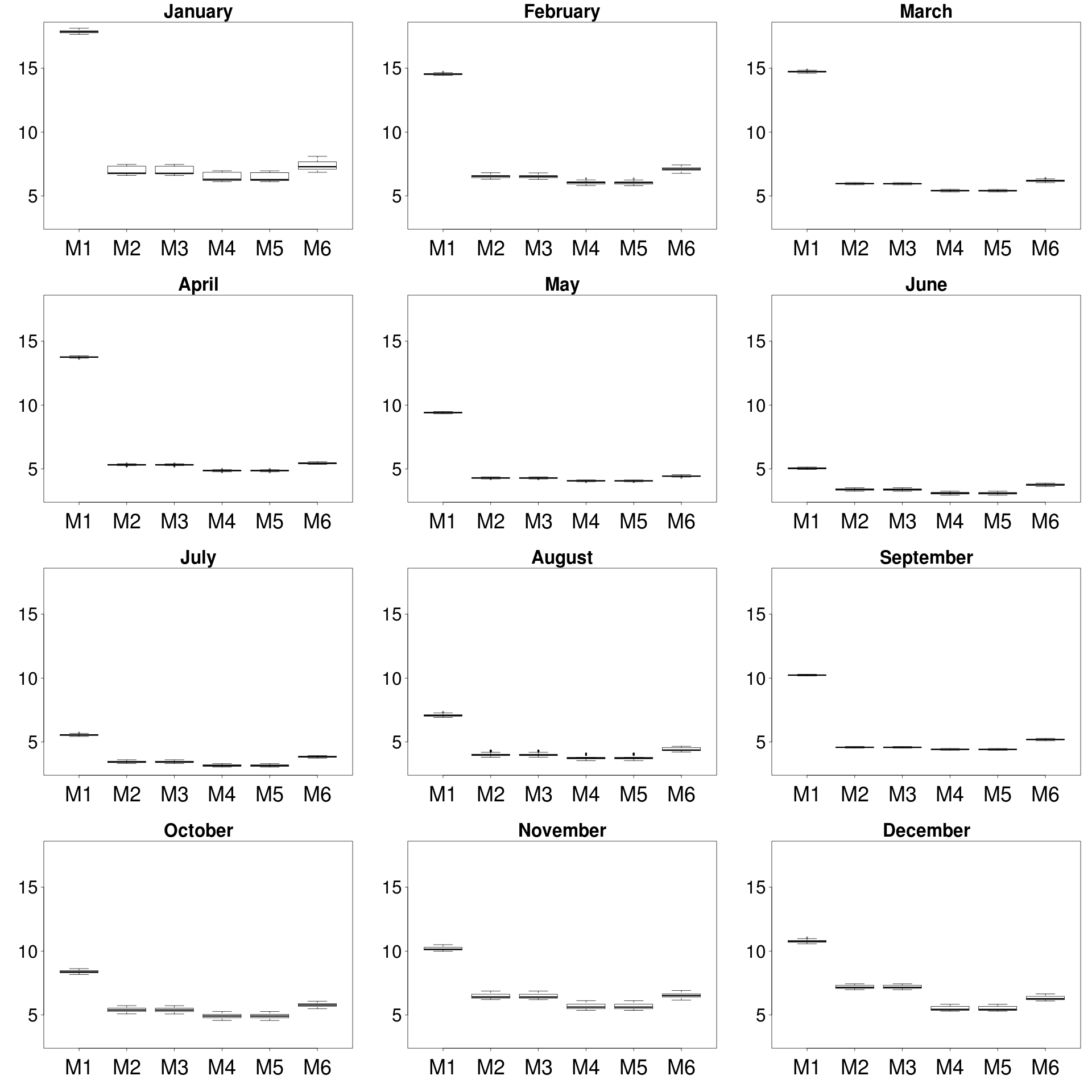}
\caption{Boxplots of root mean-squared prediction errors in various months for six different methods: (M1) AirBox data only with no calibration; (M2) AirBox data only with a global calibration procedure;
(M3) A combination of AirBox and EPA data with a global calibration procedure; (M4)  AirBox data only with the proposed spatially adaptive calibration method;
(M5) A combination of AirBox and EPA with the proposed spatially adaptive calibration method, (M6) EPA data only.}
\label{fig:rmspe}
\end{figure}

We repeated the whole procedure by randomly decomposing EPA data into training data and test data 100 times,
from which we obtained 100 RMSPE values for each method at each month.
The results are summarized as boxplots in Figure~\ref{fig:rmspe}, separately for each month.
Overall, the RMSPE values are larger in the winter (with usually higher PM$_{2.5}$ values) than in the summer
(with usually lower PM$_{2.5}$ values).
Method M1 with no calibration performed considerably worse than all the other methods.
Methods M2 and M3, which apply a global calibration, improved over Method M1 by about 32\% to 68\%.
But they were outperformed by the proposed M4 and M5 by about 3\% to 24\%,
showing the advantage of applying spatially adaptive calibrations.
Although Method M6 (utilizing high-quality EPA data) performed better than Method M1 (using only AirBox data with no calibration),
it was outperformed by M2 and M3 in almost all months except in December.
Interestingly, Method M2 performed almost the same as Method M3, and
Method M4 performed only less than 1\% worse than Method M5,
indicating that once the calibration curves are established,
we no longer require EPA data unless for locations very close to EPA stations.
Consequently, the AirBox network can almost replaces the EPA network after applying the proposed calibration procedure for PM$_{2.5}$ predictions.

\section{Summary}

The AirBox network, consisting of low-cost microsensors, provides a new way to monitor PM$_{2.5}$ at a high spatial resolution that is not possible by traditional monitoring stations.
However, its usefulness has been suspective
since measurements from AirBoxes are not accurate with many outliers, have high variability,
and are highly affected by local environmental conditions.
This paper develops an effective calibration procedure to relieve the concern.
Unlike commonly used calibration techniques, our method does not require putting AirBoxes side by side at monitoring stations.
It automatically considers local environmental conditions by borrowing information from nearby monitoring stations using a spatial varying-coefficients model.
In addition, the proposed method accounts for various aspects of data all in a regression framework, which is easy to understand and implement.

Although our model uses many parameters to describe high complexity in the dataset due to highly complex topographical conditions,
climate patterns, and pollution source distributions in Taiwan,
the complete calibration procedure is fast.
Once the calibrated model is established, we can calibrate all AirBoxes and obtain a PM$_{2.5}$ predicted map in real-time,
even if some AirBoxes are newly added to the network.
In addition, the calibration procedure is not necessary to implement frequently.
It requires to update at most weekly (or monthly) or if some significant changes in PM$_{2.5}$ chemical compositions occur somewhere.

More and more AirBoxes keep adding to the network, enabling us to obtain the PM$_{2.5}$ map at a higher spatial resolution. 
Consequently, the AirBox network has great potential to detect new emission sources and help government agencies to make proper control strategies.
It is an exciting but challenging problem.
For example, it is required to distinguish emission contributions from existing sources and outlying measurements.
We consider it an important topic for future research.

\section*{References}

\begin{description}
\item Huang, G., Chen, L.-J., Hwang, W.-H., Tzeng, S.~and Huang, H.-C.~(2018).
  Real-time PM2.5 mapping and anomaly detection from AirBoxes in Taiwan, \textit{Environmetrics},
  \textbf{29}, \url{https://doi.org/10.1002/env.2537}
\item Huber, P.~J.~and Ronchetti, E.~M.~(2009). \textit{Robust Statistics}, 2nd edition, Wiley, New York.
\item Rousseeuw, P.~J.~and van Driessen, K.~(1999). A fast algorithm for the minimum covariance determinant estimator,
  \textit{Technometrics}, \textbf{41}, 212--223.
\item Tzeng, S.~and Huang, H.-C.~(2018). Resolution adaptive fixed rank kriging, \textit{Technometrics}, \textbf{60}, 198--208.
\item Vaida, F.~and Blanchard, S.~(2005). Conditional Akaike information for mixed-effects models, \textit{Biometrika},
  \textbf{92}, 351--370.
\end{description}

\end{document}